%% file: main.tex
\documentclass[aps,prd,reprint,amsmath,amssymb,nofootinbib,longbibliography,superscriptaddress,floatfix]{revtex4-2}
\usepackage[T1]{fontenc}
\usepackage{graphicx,bm,booktabs,xcolor,array,microtype,etoolbox}
\usepackage{quantikz}
\usepackage[colorlinks=true,citecolor=blue,linkcolor=blue,urlcolor=blue]{hyperref}
\hypersetup{pdftitle={Qubit-qumode simulation of molecular torsion in photoisomerization},pdfauthor={Dimitrios Athanasakos, Florian Hechenberger, Tommaso Rainaldi, Felix Ringer, George Siopsis}}

\graphicspath{{figures/}}
\DeclareRobustCommand{\ket}[1]{\lvert #1\rangle}
\DeclareRobustCommand{\bra}[1]{\langle #1\rvert}
\newcommand{\Tr}{\operatorname{Tr}}
\newcommand{\ii}{i}
\newcommand{\e}{\mathrm{e}}
\newcommand{\R}{\mathbb{R}}
\newcommand{\Sone}{\mathbb{S}^{1}}
\newcommand{\dd}{\mathrm{d}}
\newcommand{\one}{\hat{\mathbb{I}}}

\AtBeginEnvironment{thebibliography}{\interlinepenalty=500\relax}
\makeatletter
\let\papertocsubsection\l@subsection
\renewcommand{\l@subsection}[2]{%
  \ifnum\c@tocdepth>1\relax\papertocsubsection{#1}{#2}\fi}
\let\l@subsubsection\@gobbletwo
\makeatother
\begin{document}
\title{Qubit-qumode simulation of molecular torsion in photoisomerization}
\author{Dimitrios Athanasakos}
\email{dimitrios.athanasakos@stonybrook.edu}
\affiliation{Department of Physics and Astronomy, Stony Brook University,
Stony Brook, NY 11794, USA}
\author{Florian Hechenberger}
\email{florian.hechenberger@oeaw.ac.at}
\affiliation{Marietta Blau Institute for Particle Physics, Austrian Academy of Sciences, Vienna, 1010, Austria}
\author{Tommaso Rainaldi}
\email{tommaso.rainaldi@stonybrook.edu}
\affiliation{Department of Physics and Astronomy, Stony Brook University,
Stony Brook, NY 11794, USA}
\author{Felix Ringer}
\email{felix.ringer@stonybrook.edu}
\affiliation{Department of Physics and Astronomy, Stony Brook University,
Stony Brook, NY 11794, USA}
\author{George Siopsis}
\email{siopsis@tennessee.edu}
\affiliation{Department of Physics and Astronomy, The University of Tennessee, Knoxville, TN 37996, USA}
\begin{abstract}
Photochemical reactions driven by large nuclear motion, such as the photoisomerization of retinal, depend strongly on both the global anharmonic forces of the potential and the periodic topology of molecular torsion. In this work, we develop a hybrid discrete (qubit) and continuous variable (qumode) quantum simulation framework for photochemical reactions that maps electronic states to qubits, vibrations to qumodes, and the periodic torsion to a compact qumode encoding. Applying this approach to the two-mode Hahn-Stock and three-mode PSB3 models of retinal photoisomerization, we compare i) local vibronic approximations, ii) the full trigonometric potential on the noncompact line, and iii) the same potential on the compact circle, which matches the molecular torsion topology. We numerically show that the topological description can significantly alter relevant observables. We realize the compact torsion via both a wrapped Gottesman-Kitaev-Preskill (GKP) encoding and a two-qumode circle encoding and introduce two methods that reduce the hardware resources of both encodings. Spreading an encoded coordinate over several qumodes lowers the average occupation number of each mode, and adding counterterms to the Hamiltonian reduces the required squeezing amount. We expect that these methods make percent-level simulations of the considered molecules promising targets for hybrid qubit-qumode hardware. For more complex molecules, each additional nuclear coordinate is represented directly as a single qumode or an encoded register using the same native operations, so the framework extends naturally to larger systems.

\end{abstract}
\maketitle
\tableofcontents
\input{sections/introduction}
\input{sections/model}

\input{sections/encodings}
\input{sections/results}
\input{sections/psb3}
\input{sections/resources}
\input{sections/conclusions}
\addtocontents{toc}{\protect\setcounter{tocdepth}{1}}
\input{sections/appendices}

\bibliography{refs}
\end{document}

%% file: sections/introduction.tex
\section{Introduction}
\label{sec:introduction}

In photochemistry, light absorption prepares an excited molecule that can explore configurations far from equilibrium before forming a product or returning to its initial structure. Predicting this evolution connects molecular interactions to reaction pathways, branching fractions and the time scales of chemical change. In photoisomerization, the electronic excitation drives a large-amplitude rearrangement of molecular geometry, with applications ranging from reversible molecular photoswitches to the primary photochemical event in vision~\cite{domcke2012,polli2010,Schultz:2003Azobenzene}.

Near conical intersections, where electronic potential-energy surfaces become degenerate, nonadiabatic couplings between electrons and nuclei govern the reaction. A nuclear wavepacket can transfer between electronic states, split into different branches and accumulate phases that affect its subsequent evolution~\cite{Yarkony:1996zz,worth2004,annurev:/content/journals/10.1146/annurev.physchem.57.032905.104612,domcke2012}.  Retinal photoisomerization, the rotation from the cis to the trans geometry about a carbon-carbon double bond,  illustrates this connection between electronic transitions and structural change, with femtosecond spectroscopy establishing ultrafast photoproduct formation \cite{schoenlein1991} and passage through a conical intersection~\cite{polli2010}.

Among classical methods, trajectory approaches combine electronic evolution with approximate nuclear motion~\cite{tully1990,crespootero2018}, while wavepacket methods, such as the multiconfiguration time-dependent Hartree (MCTDH) algorithm~\cite{beck2000},  retain the correlated quantum dynamics of selected nuclear coordinates. Multilayer MCTDH~\cite{vendrell2011} and tensor-network methods~\cite{greene2017,Baiardi:2019mir,xie2019} extend wavepacket propagation to tens and, in favorable cases, hundreds of modes. However, as the number of active coordinates grows, resolving their correlations and maintaining an accurate global description of the potential becomes increasingly demanding. Reduced vibronic models provide a controlled setting in which to study these questions and to compare simulation methods~\cite{koppel1984,Cigrang:2025ykc}. Reduced models of retinal range from the two-mode Hahn-Stock model~\cite{hahnstock2000,hahnstock2000cp} to the three-mode PSB3 model~\cite{marsili2019,marsili2020}, which retain the torsional coordinate and conical intersections that drive photoisomerization. For a large-amplitude reaction, even a small model must retain the features of the coordinate that govern its global motion. A molecular torsion is a continuous compact coordinate on the circle: a full rotation returns to the same molecular geometry, while a local displacement or an oscillator quadrature is continuous and noncompact, so distinct points remain distinct however far apart they lie. As we show in this work, this topological distinction alters the reaction dynamics even when the local forces are reproduced exactly. 

\begin{figure*}[t]
\centering
\includegraphics[width=\textwidth,height=0.34\textheight,keepaspectratio]{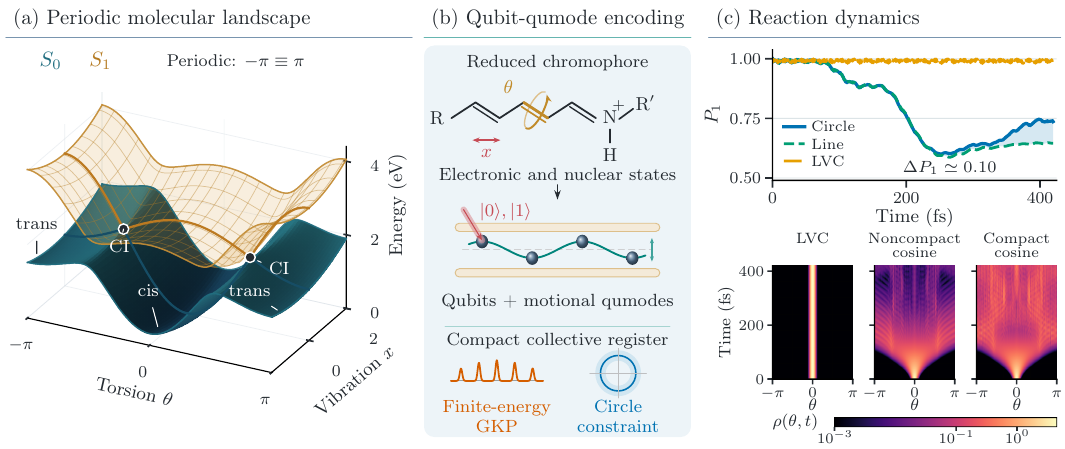}
\caption{Quantum simulation of torsional photochemistry.
(a) The Hahn-Stock adiabatic surfaces $S_0$ and $S_1$ over one full rotation, indicating the cis and trans geometries and the two conical intersections. (b) Electronic and nuclear degrees of freedom are represented by qubits and qumodes. The illustration shows trapped ion collective motional modes as qumodes. The compact torsion can be encoded through a finite-energy Gottesman-Kitaev-Preskill (GKP) construction or a circle constraint. (c) The upper panel compares electronic populations for LVC and the full trigonometric potential on the continuous noncompact line and the physical compact circle. The lower panels show the total torsional probability density $\rho(\theta,t)$, summed over electronic states and integrated over the vibration, for LVC, noncompact cosine and compact cosine, from left to right. The color scale is logarithmic. We show the noncompact density only in the central cell $[-\pi,\pi]$, so the probability that has moved beyond $\pm\pi$ does not contribute.}
\label{fig:overview}
\end{figure*}

Quantum simulation offers an alternative route to retaining the electronic and nuclear degrees of freedom and their correlations \cite{Kassal:2010xwg,Ollitrault:2020nqs,Magann:2021jyd}. Digital qubit algorithms discretize nuclear coordinates on grids or in truncated vibrational bases, with the Hamiltonian structure determining the circuit cost  \cite{Zalka:1996st,Eklund:2026sle,Kassal:2008rnk,Macridin:2018gdw,Sawaya:2020bky,Ollitrault:2020kms}.
Recent block encodings of vibrational Hamiltonians for fault-tolerant energy estimation expand the potential about the equilibrium geometry and leave large-amplitude torsion outside their scope~\cite{Mehendale:2026uxf}. Hybrid qubit-qumode architectures offer a more direct physical representation: qubits encode electronic states and qumodes, quantum harmonic oscillators with continuous canonical quadratures,  encode nuclear motion~\cite{Braunstein:2005zz,Weedbrook:2011wxo, MacDonell:2021xwm}. 
This representation makes nuclear displacements and harmonic evolution native operations. Bosonic devices have been used for vibronic spectroscopy through boson sampling~\cite{Huh:2014cli}, quantum optical emulation of molecular spectra with trapped ions~\cite{Shen:2017txh}, photonic simulation of vibrational dynamics~\cite{Sparrow:2018rjt}, anharmonic vibrational dynamics on bosonic devices~\cite{Malpathak:2024weo} and time-domain simulation of vibronic spectra with a single trapped ion~\cite{MacDonell:2022fdy}.

Coupling to a two-level system enables the preparation of arbitrary field states~\cite{Law:1996yvc} and universal control of a cavity qumode~\cite{Krastanov:2015rkt,Eickbusch:2021uod}. Neutral atoms in optical tweezers couple electronic and motional degrees of freedom~\cite{Shaw:2023apj,Hwang:2025vgp}. Trapped ions combine coherent control of internal states and collective motional qumodes~\cite{Leibfried:2003zz,Araz:2024dcy} and have simulated conical intersections, their geometric-phase interference and the chemical dynamics of vibronic Hamiltonians~\cite{Whitlow:2022oti,Valahu:2022kfg,Navickas:2024pmm}.
Gaussian operations generate Hamiltonians at most quadratic in the quadratures, so universal continuous-variable dynamics requires a non-Gaussian resource~\cite{Lloyd:1998jk,Weedbrook:2011wxo, Liu:2024mbr}. Qubit-assisted signal processing provides constructions for anharmonic vibronic potentials~\cite{Hong:2025dct}. Trigonometric gate constructions provide direct access to periodic potentials and multimode couplings through programmable Fourier synthesis~\cite{Rainaldi:2025ymn,Chalermpusitarak:2025cod} and have been realized in trapped ions together with programmable anharmonic dynamics~\cite{Rainaldi:2026wpt,McGarry:2026mkk}. Their application to photochemistry raises the central question of how to represent the compact torsional coordinate on which those periodic interactions act.

We give an overview of the problem and our approach in Fig.~\ref{fig:overview} using the Hahn-Stock model of photoisomerization, connecting the molecular potential-energy surfaces and conical intersections to a hybrid qubit-qumode representation and to reaction observables that separate local, noncompact and compact descriptions. Specifically, we distinguish three descriptions throughout this work:
\begin{enumerate}
  \renewcommand{\labelenumi}{(\roman{enumi})}
  \item Local models. Linear and quadratic vibronic coupling models, denoted LVC and QVC, expand the potential about a reference geometry and treat the torsional displacement as a coordinate on a line.
  \item Trigonometric line. The full trigonometric potential on a noncompact coordinate restores the global forces but treat $\theta$ and $\theta+2\pi$ as distinct configurations. 
  \item Trigonometric circle. The same potential is placed on the circle, the physical configuration space of the torsion, where $\theta$ and $\theta+2\pi$ coincide.
\end{enumerate}
Comparing (i) with (ii) tests the approximation to the potential and comparing (ii) with (iii) tests the topology of the configuration space. The true model requires (iii) to correctly describe the torsion, but (ii) is the cheaper option since qumodes are naturally noncompact. It can be accurate until amplitudes that explore different angular cells produce an observable difference.

On the circle, amplitudes that reach the same molecular geometry after different winding histories can interfere. On the line, those amplitudes remain in separate translated cells, so electronic populations and product probabilities potentially differ despite identical periodic forces. We illustrate the two physical effects in Fig.~\ref{fig:overview}(c): local confinement can suppress torsional motion, while the full-potential line and circle track each other initially before separating at later times as winding interference alters electronic populations and product probabilities.

Continuous compact variables also occur in lattice gauge theory, where an angular field and its conjugate electric field obey the same rotor algebra as molecular torsion. Hybrid qubit-qumode proposals~\cite{Marshall:2015mna,Davoudi:2021ney,Jha:2023ecu,Crane:2024tlj,Briceno:2023xcm,Ale:2024uxf,Araz:2024dcy, Liu:2024mbr,Ale:2025sxz,Abel:2025zxb, Ale:2026hqe,Shinbrough:2026dyu,Wang:2026kcu} and experiments~\cite{Than:2025gso} have motivated two bosonic encodings that distinguish this algebra from that of a noncompact qumode. A Gottesman-Kitaev-Preskill (GKP) translation stabilizer \cite{Gottesman:2000di} selects a compact rotor sector within the Bloch decomposition of a qumode~\cite{Ale:2026hqe}. A second construction constrains two qumode coordinates to a circle and uses their angular motion for the compact variable~\cite{Ale:2025sxz}.
In the molecular setting of this work, these two constructions must preserve nonadiabatic transfer and reaction observables while coupling the torsion to additional nuclear degrees of freedom. We develop and test both realizations under those requirements.

Our primary application is the two-mode Hahn-Stock model of retinal photoisomerization~\cite{hahnstock2000,hahnstock2000cp}, which has the form of the photoisomerization models introduced by Seidner and Domcke~\cite{seidner1994}. They solved these models in a direct-product basis of diabatic electronic states, free-rotor states and harmonic-oscillator states, the classical counterpart of the qubit, rotor and qumode registers used here. Its two electronic states couple to the torsional reaction coordinate and one vibration, giving conical intersections and transfer between cis ($\theta=0$) and trans ($\theta=\pi$) regions in a model that admits accurate reference propagation. Extensions with further vibrational modes have also served as benchmarks for quantum and quantum-classical dynamics~\cite{pereira2023}. Versions of the model with additional bath modes have served as benchmarks for tensor-network and open-system methods~\cite{Lyu:2022dmn,schile2019}. We compare the local vibronic models, the full trigonometric potential on the line, and the physical compact model on the circle through their effects on population transfer and on the time-dependent branching between cis and trans basins. Additionally, we study the three-mode PSB3 model
\cite{marsili2019,marsili2020}, which provides a more detailed chromophore application, with additional mode coupling, coherent motion and nuclear recurrences
\cite{aarabi2025}.

We map each degree of freedom to the resource that describes it most directly. The electronic states naturally map onto a qubit and the coupling vibration, a noncompact continuous variable, maps directly onto a qumode. The compact torsion requires more care because it is not native. For that we adopt two bosonic encodings: a wrapped GKP-style encoding on a qumode or a circle embedded in two qumodes. In practice, the states corresponding to these two encodings are replaced by finite-energy GKP states and finite-width circle states. We quantify the Fock-space occupation each encoding requires and show how to exploit two freedoms that make the encodings cheaper without changing the simulated dynamics. Passive mode mixing~\cite{Reck:1994ujl,Weedbrook:2011wxo} distributes an encoding over several physical qumodes, conserving the total excitation number while lowering the occupation of each mode. A similar uniform redistribution fixes the local cutoff in qubit simulations of boson sampling~\cite{Oh:2026ljm}. Second, counterterms added to the Hamiltonian, leave the encoded dynamics unchanged and reduce the sensitivity of the dynamics to the finite width of the encoding. This relaxes the required squeezing and lowers the total occupation that redistribution then shares among the modes.

We also build the time step from elementary gates and compare it with a qubit implementation of the same Hamiltonian, in the broader context of bosonic encodings, planar rotors and vibronic simulation~\cite{Sawaya:2020bky,Moeed:2025wtt,Nourse:2025qus}. Finally, we estimate the resources these simulations need under stated assumptions. 
The representation also extends to larger molecules. Each additional vibration adds one qumode and each additional torsion one encoded register, with the same native operations. While classical wavepacket methods, such as MCTDH, can treat vibronic models with about two dozen modes, such as the 24-mode model of pyrazine~\cite{raab1999,beck2000}, their cost grows exponentially with the number of strongly correlated coordinates~\cite{vendrell2011}, and for tensor networks with the entanglement that builds up during the dynamics~\cite{Baiardi:2019mir}. Resource estimates for vibronic dynamics on quantum hardware~\cite{Motlagh:2024yhh,Kang:2023xfb} indicate where quantum simulations can extend beyond this regime. 

The rest of the paper is organized as follows. In Sec.~\ref{sec:model} we introduce the Hahn-Stock Hamiltonian and relevant observables. In Sec.~\ref{sec:encodings} we develop the compact encodings, their extension over several qumodes and the Hamiltonian counterterms. In Sec.~\ref{sec:results} we present the Hahn-Stock dynamics and finite-resource benchmarks and in Sec.~\ref{sec:psb3} we develop the PSB3 application. In Sec.~\ref{sec:resources} we compare qubit and qumode requirements and discuss the experimental tradeoffs of the two encodings.

%% file: sections/model.tex
\section{Molecular torsion and the Hahn-Stock model}
\label{sec:model}

\subsection{A qubit coupled to a rotor and a vibration}

The minimal Hahn-Stock model of photoisomerization couples two electronic states (a qubit) to two active nuclear coordinates: a torsion $\theta$ (a planar rotor) and a coupling vibration $x$ (a harmonic oscillator)~\cite{hahnstock2000,hahnstock2000cp}. Its reduced configuration space retains the two ingredients that are central to the present work. Rotations along $\theta$ change the electronic energy landscape globally, while the vibration $x$ permits nonadiabatic population transfer.

We work in the diabatic electronic basis $\{\ket{0}, \ \ket{1}\}$, whose states are independent of the nuclear coordinates and are chosen such that the potential matrix is a smooth function of them. In this basis, the nuclear kinetic terms are diagonal and the electronic coupling is carried entirely by the potential. The physical torsion $\theta$ is a compact coordinate on the circle $\Sone$ and the vibration $x$ is described by a dimensionless quadrature on $\R$ with $[\hat x,\hat p_x]=\ii$. Therefore, the molecular Hilbert space is
\begin{equation}
 \mathcal H_{\rm HS}=\mathbb C^2\otimes L^2(\Sone)
 \otimes L^2(\R).
 \label{eq:molspace}
\end{equation}
For simplicity, we use $\hbar=1$ in operator expressions and quote molecular energies in
eV and times in fs, unless otherwise stated. The molecular Hamiltonian is
\begin{align}
 \hat H_{\rm HS} &= \frac{G}{2}\hat L^2+
 \frac{\omega}{2}(\hat p_x^2+\hat x^2)+\hat V(\hat\theta,\hat x),
 \label{eq:hs}\\
 V(\theta,x)&=
 \begin{pmatrix}
 \frac{W_0}{2}(1-\cos\theta)&\lambda x\\
 \lambda x&E_1-\frac{W_1}{2}(1-\cos\theta)+\kappa x
 \end{pmatrix}.
 \label{eq:hsV}
\end{align}
Here $V(\theta,x)$ is the coordinate-space matrix of the potential operator $\hat V$ in the $\{\ket{0}, \ \ket{1}\}$ basis while the kinetic terms act as the identity in the electronic space. The parameters of this model are $(E_1,W_0,W_1,\omega,\kappa,\lambda) =(2.48,3.60,1.09,0.19,0.10,0.19)$ eV and the inverse torsional moment of inertia $G=4.84\times10^{-4}$ eV. This two-coordinate parametrization is also implemented in QuantumModelLib~\cite{quantummodellib}. Because the cosine terms carry opposite signs in the two diabats, the cis geometry ($\theta=0$) is a minimum of the ground diabat $\ket{0}$ and a maximum of the excited diabat $\ket{1}$. The diagonal coupling $\kappa x$ displaces the excited potential along the vibration, while $\lambda x$ drives electronic transitions.

On $\Sone$, periodic boundary conditions restrict the angular momentum operator $\hat L=-\ii\partial_\theta$ to integer eigenvalues. Together with the unitary shift operator $\hat U=e^{i \hat\theta}$, the rotor algebra  reads
\begin{equation}
  [\hat L,\hat U]=\hat U,\qquad
 \hat L\ket{\ell}=\ell\ket{\ell},\quad \ell\in\mathbb Z.
 \label{eq:rotoralgebra}
\end{equation}

If $\Sone$ is unwrapped onto the real line $\R$, the operators $\hat L=-\ii\partial_\theta$ and $\hat U=\e^{\ii\hat\theta}$ satisfy the same commutation relation $[\hat L,\hat U]=\hat U$, but $\hat L$ acquires a continuous spectrum on $\R$. Although wavepackets on $\Sone$ and $\R$ experience identical local forces under Eq.~\eqref{eq:hsV}, their global boundary conditions differ. Trajectories that wind around the torsion by different multiples of $2\pi$ recombine and interfere on $\Sone$ while on $\R$ they propagate into distinct spatial cells.
To make this explicit, let $\psi_m(\theta,x,t)$ be the two-component line wavefunction in the cell at $\theta+2\pi m$, with the line packet initially in the central cell and its coherent periodization as the circle state. The compact amplitude is then the sum over translated line amplitudes. The
two reduced angular densities are
\begin{align}
 p_{\rm circ}(\theta,t)&=\int\dd x\,
       \left\|\sum_m\psi_m(\theta,x,t)\right\|^2,\nonumber\\
 p_{\rm line}^{\rm fold}(\theta,t)&=\int\dd x\,
       \sum_m\|\psi_m(\theta,x,t)\|^2.
 \label{eq:windinginterference}
\end{align}
Compactness retains the cross terms between winding histories, while the folded line readout omits them. The remaining degrees of freedom determine which cross terms survive the partial trace. If two winding contributions factorize as $\psi_m=c_m\xi_m$, their cross term is proportional to $c_mc_{m'}^*\langle\xi_{m'}|\xi_m\rangle$. Orthogonal vibrational states distinguish the paths and suppress their interference, while identical ones preserve it. At early times the winding amplitudes are small, so the line reproduces the compact observables until they grow.

We quantify the effect of this topological distinction on the reaction observables in Sec.~\ref{sec:results}.

\subsection{Potential-energy surfaces and conical intersections}
\label{sec:surfaces}

In this work we will simulate the dynamics of the Hahn-Stock model in the diabatic $\{\ket{0},\ket{1}\}$ basis, but the adiabatic potential-energy surfaces make the reaction pathways transparent: the descent from the cis geometry, the branching at the conical intersections and the arrival in the trans basin or the return to the cis geometry. Denoting the Pauli matrices by
$\hat\sigma_i$ with $i\in\lbrace x, y, z\rbrace$, we write
\begin{align}
 \hat V&=\bar v(\hat\theta,\hat x)+d_z(\hat\theta,\hat x)\hat\sigma_z\nonumber +\lambda\hat x\hat\sigma_x,\nonumber\\
 \bar v(\theta,x)&=\frac{E_1}{2}+\frac{W_0-W_1}{4}(1-\cos\theta)
             +\frac{\kappa x}{2},\nonumber\\
 d_z(\theta,x)&=\frac{W_0+W_1}{4}(1-\cos\theta)-\frac{E_1}{2}
             -\frac{\kappa x}{2}.
 \label{eq:paulipotential}
\end{align}
Diagonalizing the electronic structure at fixed nuclear coordinates yields the $2\pi$-periodic adiabatic potential-energy surfaces $S_0(\theta,x)$ and $S_1(\theta,x)$ shown in Fig.~\ref{fig:overview}(a):

\begin{equation}
 S_{0,1}(\theta,x)=\frac{\omega x^2}{2}+\bar v(\theta,x)
             \mp\sqrt{d_z(\theta,x)^2+\lambda^2x^2}.
 \label{eq:adiabaticsurfaces}
\end{equation}
The adiabatic gap closes when $d_z(\theta,x)$ and $\lambda x$ vanish simultaneously. For  $\lambda\ne0$, the intersections occur at $x=0$ and
\begin{equation}
 \cos\theta_c=1-\frac{2E_1}{W_0+W_1},
 \label{eq:ci}
\end{equation}
giving $\theta=\pm\theta_c$, where the subscript $c$ labels the conical intersection and $\theta_c=93.30^\circ$. The two nuclear coordinates resolve the conical geometry around each crossing. For example, near the intersection at $\theta=+\theta_c$, let $\delta\theta=\theta-\theta_c$ and
$a_c=(W_0+W_1)\sin\theta_c/4$. The leading gap is

\begin{equation}
 S_1-S_0=2\sqrt{(a_c\delta\theta-\kappa x/2)^2+\lambda^2x^2}
 +O(\delta\theta^2,x^2).
 \label{eq:branchingplane}
\end{equation}
The tuning coordinate $a_c\delta\theta-\kappa x/2$ and the coupling coordinate $\lambda x$ linearly span the branching plane of the intersection.

The energetic ordering of the diabatic states $\bra{0}\hat V\ket{0}$ and $\bra{1}\hat V\ket{1}$ flips across the intersection.  At the cis geometry $(\theta,x)=(0,0)$ their energies are equal to $0$ and $E_1=2.48$ eV, while at the trans geometry $(\theta,x)=(\pi,0)$ they are equal to $W_0=3.60$ eV and $E_1-W_1=1.39$ eV, respectively. Therefore, the lower adiabatic surface $S_0$ is dominated by $\ket{0}$ near cis and by $\ket{1}$ near trans.

\subsection{Photoexcitation and molecular scales}
\label{sec:preparation}

Before photoexcitation, the molecule occupies the nuclear ground state of the $\ket{0}$ diabat, $\ket{0}\otimes\ket{\chi_{\rm cis}}\otimes\ket{n{=}0}_x$. The vibrational factor $\ket{n{=}0}_x$ is the zero Fock state associated with the vibrational degree of freedom and $\ket{\chi_{\rm cis}}$ is the ground state of $G\hat L^2/2+W_0(1-\cos\hat\theta)/2$ on the circle.
Expanding around $\theta = 0$ (the cis minimum) we obtain the harmonic oscillator approximation
\begin{equation}
 \frac{W_0}{2}(1-\cos\theta)=\frac{W_0}{4}\,\theta^2+O(\theta^4).
 \label{eq:ciswell}
\end{equation}
Its frequency and ground-state position width are
\begin{align}
 \Omega_\theta&=\sqrt{\frac{GW_0}{2}}=0.0295\,\text{eV},\nonumber\\
 \sigma_{\theta,0}&=\left(\frac{G}{2W_0}\right)^{1/4}=0.0905\,\text{rad}.
 \label{eq:initialscales}
\end{align}
Because $\sigma_{\theta,0}\ll \pi$,  the initial packet is tightly localized within the cis basin where the harmonic approximation is accurate.

Photoexcitation enters as a sudden electronic transition at fixed nuclear coordinates, $\ket{1}\bra{0}\otimes\one$. This description is valid for pulses much shorter than the two characteristic nuclear periods, $2\pi\hbar/\omega=21.8$ fs for the coupling vibration  and $2\pi\hbar/\Omega_\theta=140$ fs for the torsional well.  The resulting photoexcited state is the initial state of the dynamics
\begin{equation}
 \ket{\Psi(0)}=\ket{1}\otimes\ket{\chi_{\rm cis}}
 \otimes\ket{n{=}0}_x .
 \label{eq:initial}
\end{equation}
This specifies a product Franck-Condon preparation, which deposits in the molecule the vertical excitation energy
\begin{equation}
 E_{\rm vert}=\Big[\bra{1}\hat V\ket{1}-\bra{0}\hat V\ket{0}
 \Big]_{(\theta,x)=(0,0)}=E_1 .
 \label{eq:vertical}
\end{equation}

At cis, the excited diabat is at a local maximum of the torsional potential, so the wavepacket accelerates away from the cis region toward the intersection region. If the entire excitation energy $E_1$ is converted into torsional rotation, $G\ell_E^2/2=E_1$, the angular momentum reaches its energetic bound $\ell_E$ corresponding to an angular velocity $G\ell_E/\hbar$ and rotation period $T_E$:
\begin{equation}
 \ell_E=\sqrt{\frac{2E_1}{G}}\simeq101,\qquad
 T_E=\frac{2\pi\hbar}{G\ell_E}\simeq84.4\ {\rm fs}.
 \label{eq:energyscales}
\end{equation}
 A packet initially $0.09$ rad wide spreads over angular momenta up to $\ell\sim |\ell_E|$, requiring the rotor register to resolve roughly $200$ angular momentum states. A propagation window of several hundred femtoseconds covers several full windings at $T_E\simeq84$ fs and ten to twenty vibrational periods, so it probes the global configuration space. For simulations on $\R$, the localized packet is initialized in the central cell $[-\pi, \pi]$.

\subsection{Local vibronic approximations}
\label{sec:local}

Standard vibronic models expand the diabatic potential around a reference geometry~\cite{koppel1984,domcke2012} on the line $\R$ with momentum $\hat p_\theta=-\ii\partial_\theta$. The linear vibronic coupling  (LVC) approximation retains a state-independent  potential along with linear  displacements.
\begin{align}
 \hat H_{\rm LVC}&=\frac{G}{2}\hat p_\theta^2+
 \frac{W_0}{4}\hat\theta^2+
 \frac{\omega}{2}(\hat p_x^2+\hat x^2)\nonumber\\
 &\quad +(E_1+\kappa\hat x)\,\frac{\one_2-\hat\sigma_z}{2}
 +\lambda\hat x\hat\sigma_x .
 \label{eq:lvc}
\end{align}
In Eq.~\eqref{eq:lvc} the torsion appears only through its own harmonic Hamiltonian, which commutes with every other term. The torsion therefore decouples and the electronic transitions are only driven by $x$ alone, while the tails of the localized torsional state give an exponentially suppressed probability in the trans region.

The quadratic vibronic coupling (QVC) model restores the state-dependent quadratic torsional curvature,
\begin{align}
 \hat H_{\rm QVC}&=\frac{G}{2}\hat p_\theta^2+
 \frac{\omega}{2}(\hat p_x^2+\hat x^2)\nonumber\\
 &\quad+
 \begin{pmatrix}
 W_0\hat\theta^2/4&\lambda\hat x\\
 \lambda\hat x&E_1-W_1\hat\theta^2/4+\kappa\hat x
 \end{pmatrix}.
 \label{eq:qvc}
\end{align}
While $\hat H_{\rm QVC}$ captures the initial  acceleration and the first nonadiabatic passage, its excited diabatic surface is an inverted parabola that is unbounded from below, lacking the global restoring forces of the cosine potential.

\subsection{Observables and reference calculations}
\label{sec:observables}

We characterize the time-evolved state $\ket{\Psi(t)}=\e^{-\ii\hat Ht/\hbar}\ket{\Psi(0)}$ via expectation values $\langle\hat O\rangle_t=\bra{\Psi(t)}\hat O\ket{\Psi(t)}$. For this paper, we mainly focus on three physical questions. Whether electronic population has transferred, whether the torsion has reached the trans basin and whether a photoproduct has formed, that is, whether the torsion is in the trans basin while the electronic state is on the lower adiabatic surface $S_0$.

To calculate the electronic population transfer, we define the excited-configuration population
\begin{equation}
 P_1(t)=\big\langle\ket{1}\bra{1}\otimes\one_\theta\otimes\one_x\big\rangle_t
 =\Tr\big[\hat\rho_{\rm el}(t)\ket{1}\bra{1}\big],
 \label{eq:p1}
\end{equation}
where $\one_\theta$ and $\one_x$ are the identities on the torsional and vibrational factors and $\hat\rho_{\rm el}(t)=\Tr_{\theta,x}\ket{\Psi(t)}\bra{\Psi(t)}$ is the reduced electronic density matrix obtained by tracing out those nuclear degrees of freedom.

Nuclear arrival in the trans basin is defined by the trans indicator $\chi_{\rm tr}(\theta)$ for $|\theta\bmod2\pi|>\pi/2$, corresponding to the operator $\hat\chi_{\rm tr}=\one_2\otimes\chi_{\rm tr}(\hat\theta)\otimes\one_x$. Since $|\theta\bmod2\pi|>\pi/2$ is equivalent to $\cos\theta<0$, the indicator is a function of the globally defined rotor operator alone
\begin{equation}
 \chi_{\rm tr}(\hat\theta)=\Theta(-\cos\hat\theta),\qquad
 \cos\hat\theta=\frac{\hat U+\hat U^\dagger}{2},
 \label{eq:transindicator}
\end{equation}
with $\Theta$ the Heaviside step function.

Resolving chemical product formation requires projecting onto the adiabatic basis. After the vertical excitation, the packet descends  adiabatically on $S_1$. Near the conical intersection, where the gap closes, adiabatic evolution breaks down and the packet  branches. The component that flips to $S_0$ continues into the trans basin. Conversely, the branch that remains in $S_1$ faces a potential barrier rising toward $W_0=3.60$ eV at trans, above the available energy $E_1$, so it reflects back to the intersection region where it branches again. The part that further remains on $S_1$ returns toward the cis region before descending to the intersection again. Product formation at a given time therefore requires two conditions jointly: arrival in the trans basin and support on $S_0$. Each condition alone is reversible, since trans amplitude on the upper surface can swing back through the intersection and lower-surface amplitude at cis is recovered reactant. The potential matrix has the form $\bar v\,\one_2+\vec h\cdot\hat{\vec\sigma}$ with $\vec h=(\lambda x,0,d_z)$, so its eigenvalues are $\bar v\pm|\vec h|$ and its eigenvectors depend only on the direction of $\vec h$. Away from the conical intersections, where $\vec h=0$, the projector onto the lower surface is
\begin{equation}
 \hat\Pi_-(\theta,x)=\frac12\left[
 \one_2-\frac{d_z\hat\sigma_z+\lambda x\hat\sigma_x}
 {\sqrt{d_z^2+\lambda^2x^2}}\right] =\frac12\left[
 \one_2-\frac{\vec h}
 {|\vec h|}\cdot\hat{\vec\sigma}\right]  ,
 \label{eq:adiabaticprojector}
\end{equation}
which defines the finite-time product probability $Q_-$ and the
conditional product fraction $Y_-(t)$, the trans probability given
that the molecule is on the lower surface,
\begin{align}
 Q_-(t)&=\langle\hat\chi_{\rm tr}\hat\Pi_-(\hat\theta,\hat x)\rangle_t
 =P({\rm trans},S_0),\nonumber\\
 Y_-(t)&=\frac{Q_-(t)}{\langle\hat\Pi_-\rangle_t}
 =\frac{P({\rm trans},S_0)}{P(S_0)}=P({\rm trans}\,|\,S_0).
 \label{eq:yield}
\end{align}
Without an environment the dynamics is closed and $Q_-$ keeps fluctuating rather than settling to a steady-state quantum yield. The observable $Y_-$ measures the branching fraction of the $S_0$ population between trans product and cis reactant.

We also consider the return probability
\begin{equation}
 \mathcal R(t)=|\langle\Psi(0)|\Psi(t)\rangle|^2,
 \label{eq:return}
\end{equation}
defined relative to the complete initial state of each model or encoding.

Throughout the paper, we compare the curves associated with the various models and encodings to a converged reference of the observable $O$ and demand that the largest absolute error over the time window
$\mathcal T$,
\begin{equation}
 \epsilon_O=\max_{t\in\mathcal T}|O_{\rm enc}(t)-O_{\rm ref}(t)|,
 \label{eq:error}
\end{equation}
is less than a given tolerance. For most of the simulations, the tolerance we set is $0.01$. However, for experiments that only need to detect the effect of compactness a potentially larger tolerance can suffice, which we discuss in Sec.~\ref{sec:resource:occupation}. Since $Y_-$ is only well defined when the wavepacket has a non-zero weight in $S_0$ , we quote its value only at times when  $\langle\hat\Pi_-\rangle_t\ge0.05$. Numerical details can be found in Appendix~\ref{app:numerical}.

%% file: sections/encodings.tex
\section{Compact coordinates with qubits and qumodes}
\label{sec:encodings}

\begin{figure*}[t]
 \centering
 \includegraphics[width=.8\textwidth]{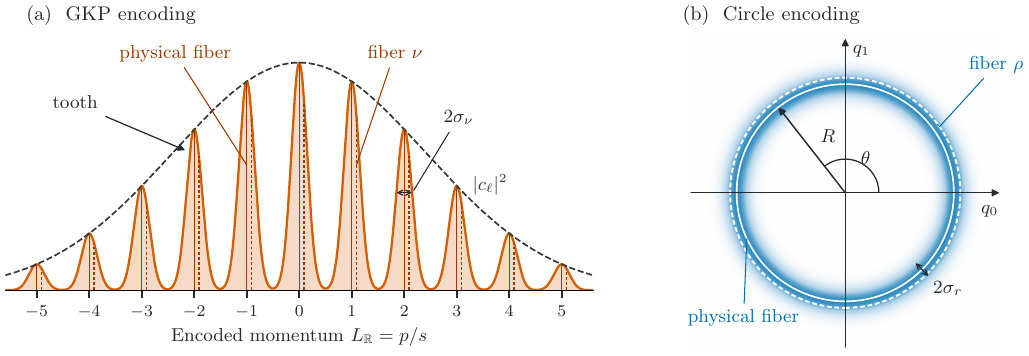}
 \caption{Finite-energy compact encodings, with widths exaggerated for
 visibility. (a) Wrapped GKP encoding: momentum distribution of a
 finite-energy comb, Eq.~\eqref{eq:teeth}. Each integer angular momentum
 $\ell$ carries a tooth of width $\sigma_\nu$ of Eq.~\eqref{eq:width}
 and the molecular coefficients $|c_\ell|^2$ set the envelope.
 (b) Two-qumode circle encoding: position distribution of a
 finite-width annulus, Eq.~\eqref{eq:ringstate}, of radius $R$ and
 radial width $\sigma_r\simeq R/(4A)$ in the plane of $q_0$ and $q_1$.
 The torsion $\theta$ is the polar angle.}
 \label{fig:encodings}
\end{figure*}

A qumode is a harmonic oscillator with the standard canonical pair $[\hat{q},\hat{p}]=i\one$ on
$L^2(\mathbb R)$, with continuous noncompact position and momentum
coordinates. Therefore, it naturally
encodes the molecular vibration. However, the torsion obeys
the rotor algebra of Eq.~\eqref{eq:rotoralgebra}, with integer angular
momentum on a compact coordinate.
The molecular potential of our model is periodic in $\hat{\theta}$ and
carries over to the qumode under $\hat{\theta}\mapsto\hat{q}$, but the angular identification of the
compact coordinate must be encoded. We
use two encodings of a rotor in qumodes. The first one wraps the qumode line onto the
circle by selecting a translation sector of one
qumode via a GKP construction~\cite{Gottesman:2000di, Ale:2026hqe}. The second one embeds the circle as a
constrained subspace of two qumodes~\cite{Ale:2025sxz}.
Figure~\ref{fig:encodings} sketches both in their finite-energy form.
Their finite-energy errors have a common description in terms of
conserved sectors, which also provides systematic ways to reduce them.
Both constructions extend to collective encodings distributed over
several qumodes and admit Hamiltonian counterterms.

\subsection{GKP encoding and the direct integral of rotors}
\label{sec:gkp}

The logic of this encoding is to make the qumode coordinate
periodic by construction. Define the unwrapped angle
$\hat{\theta}_{\mathbb R}\equiv s\hat{q}$, where $s>0$ and identify it modulo $2\pi$,
$\hat{\theta}_{\mathbb R}\simeq\hat{\theta}_{\mathbb R}+2\pi j$ with
$j\in\mathbb Z$, equivalently $\hat{q}\simeq\hat{q}+2\pi j/s$. Once the
identification holds, the rotor operators carry over through the formal
replacements
\begin{equation}
 \hat{\theta}\mapsto\hat{\theta}_{\mathbb R}=s\hat{q},\qquad
 \hat{U}=e^{i\hat{\theta}_{\mathbb R}},\qquad
 \hat{L}_{\mathbb R}=\hat{p}/s.
 \label{eq:gkpoperators}
\end{equation}
The identification itself is enforced by a stabilizer. The idea is that we only work
within the $+1$ eigenspace of the
translation
\begin{equation}
 \hat{S}=e^{2\pi i\hat{L}_{\mathbb R}},
 \label{eq:gkpstabilizer}
\end{equation}
which shifts the unwrapped angle $\hat{\theta}_{\mathbb R}$ by one
period $2\pi$. Equivalently, the stabilizer in Eq.~\eqref{eq:gkpstabilizer} acts as the identity on the space it stabilizes and that is exactly where Spec$(\hat{L}_{\mathbb{R}})\simeq \mathbb Z$, namely, the rescaled momentum has integer values. Within this sector, the most general state is
\begin{equation}
 \sum_{\ell\in\mathbb Z}c_\ell\ket{p=s\ell}.
 \label{eq:idealcomb}
\end{equation}
The above can equivalently be seen as a superposition of comb states $\ket{\theta}\equiv \sum_{\ell\in\mathbb Z} e^{-i\ell\theta}\ket{p=s\ell}$, which correspond exactly to the encoded angular states of the rotor in the qumode's Hilbert space. These belong to the GKP family. Indeed, each encoded $\ket{\theta}$ state is a so-called qunaught state, i.e. a grid state that encodes no logical information, unlike the usual GKP state that encodes a logical qubit.
Thus the GKP identification makes the momentum spectrum discrete while
retaining the full angular wavefunction.
The amplitudes $c_\ell$, which can themselves be states of the other
molecular coordinates, are determined by the molecular wavefunction. A momentum
eigenstate occupies a single tooth of the comb, while a localized angular wavepacket requires a
coherent superposition across many teeth.

In order to formally understand why we can wrap the real line onto the circle, and thus connect the rotor with the oscillator, we make use of the Zak transform. Every
eigenvalue of $\hat{L}_{\mathbb R}$ on the line splits uniquely as
$L_{\mathbb R}=\ell+\nu$ with an integer part $\ell\in\mathbb Z$ and a
fractional part $-1/2\le\nu<1/2$. The integer part plays the role of a
rotor angular momentum. The fractional part is read out by the stabilizer, since $\hat{S}$ has eigenvalue $e^{2\pi i\nu}$ on such a
state and the encoded sector of Eq.~\eqref{eq:idealcomb} corresponds to $\nu=0$.
The Zak transform organizes the whole line by this fractional momentum
and gives the unitary decomposition~\cite{Ale:2026hqe}
\begin{equation}
 L^2(\mathbb R)\simeq
 \int_{-1/2}^{1/2}{}^{\oplus}d\nu\,
 L^2_\nu(\mathbb S^1).
 \label{eq:zak}
\end{equation}
Each integrand $L^2_\nu(\mathbb S^1)$ of the direct integral is called a
fiber. Its wavefunctions satisfy the twisted periodicity condition
$\psi(\theta+2\pi)=e^{2\pi i\nu}\psi(\theta)$. Multiplying them by
$e^{-i\nu\theta}$ makes them strictly $2\pi$ periodic and on these
periodic wavefunctions the Hamiltonian of the fiber is
\begin{equation}
 \hat{H}_\nu=\frac{G}{2}(\hat{L}+\nu\one)^2+
 \frac{\omega}{2}(\hat{p}_x^2+\hat{x}^2)+\hat{V}(\hat{\theta},\hat{x}).
 \label{eq:blochsector}
\end{equation}
Here, $\hat{L}=-i\partial_\theta$ has integer eigenvalues on the periodic
fiber. These fiber Hamiltonians assemble into the line Hamiltonian
\begin{equation}
 \hat{H}_{\mathbb R}=\frac{G}{2}\hat{L}_{\mathbb R}^2+
 \frac{\omega}{2}(\hat{p}_x^2+\hat{x}^2)+\hat{V}(s\hat{q},\hat{x})
 ,
 \label{eq:linehamiltonian}
\end{equation}
which is obtained by substituting Eq.~\eqref{eq:gkpoperators} into Eq.~\eqref{eq:hs}. Since $[\hat{H}_{\mathbb R},\hat{S}]=0$, the fiber label $\nu$ is conserved and the physical compact molecular Hamiltonian is $\hat{H}_0$, i.e. the one with
vanishing fiber label, or equivalently, where the angular momentum
spectrum is integer. The periodic potential acts identically in every
fiber, so the fibers differ only through the kinetic shift visible in
Eq.~\eqref{eq:blochsector}.

This description is the ideal limit of the encoding. In reality, a single fiber is spanned by momentum eigenstates and is non-normalizable in $L^2(\mathbb R)$. A
finite-energy state carries a distribution over fiber labels concentrated near $\nu=0$. Because $\nu$ is conserved, this distribution is fixed in time and only the state within each fiber evolves.

A normalizable state follows by giving each tooth of the ideal comb a
finite width. Writing the periodic molecular state through its angular
Fourier coefficients,
$\psi_{\rm mol}(\theta,x)=\sum_\ell c_\ell(x)e^{i\ell\theta}$, we
replace the momentum delta function at each integer $\ell$ by the same
narrow Gaussian
\begin{align}
 \tilde\Psi_r(L_{\mathbb R},x)&=\mathcal N\sum_\ell
 e^{-(L_{\mathbb R}-\ell)^2/(4\sigma_\nu^2)}\,c_\ell(x),
 \label{eq:teeth}\\
 \sigma_\nu&=\frac{e^{-r_{\rm eff}}}{\sqrt{4\pi}}.
 \label{eq:width}
\end{align}
No overall envelope over $\ell$ is needed, since the coefficients
$c_\ell$ of a normalizable molecular state already decay beyond the
populated momenta. For $\sigma_\nu\ll1/2$ the teeth do not overlap, so
every fiber carries the same molecular state with the weight
$w(\nu)=e^{-\nu^2/(2\sigma_\nu^2)}/\sqrt{2\pi\sigma_\nu^2}$,
concentrated near the physical fiber $\nu=0$.

The effective squeezing $r_{\rm eff}$ sets the tooth width in rotor
units. Since $\hat{L}_{\mathbb R}=\hat{p}/s$, the physical momentum
width is $s\sigma_\nu$. Matching it to the width $e^{-r_{\rm phys}}/\sqrt2$
of a momentum-squeezed vacuum gives the physical squeezing and its
decibel value,
\begin{align}
 r_{\rm phys}&=r_{\rm eff}+\log\frac{\sqrt{2\pi}}{s},\quad
 S_{\rm dB}=\frac{20r_{\rm phys}}{\log10}.
 \label{eq:physicalwidth}
\end{align}
Because the fiber label is conserved, a measurement returns the average over the fiber distribution. Thus, for a periodic observable
$\hat{O}$, we have
\begin{equation}
 \langle\hat{O}\rangle(t)=\int_{-1/2}^{1/2}d\nu\,
 w(\nu)\langle\psi_\nu(t)|\hat{O}|\psi_\nu(t)\rangle.
 \label{eq:fiberaverage}
\end{equation}
A conditional probability such as $Y_-$ is the ratio of the averaged
numerator and denominator. The fibers differ only through the
kinetic shift of Eq.~\eqref{eq:blochsector}, so the average dephases
the rotor. For a free rotor the coherence between $\ell$ and $\ell'$
acquires the factor
\begin{equation}
 \exp\left[-\frac12G^2\sigma_\nu^2t^2(\ell-\ell')^2\right].
 \label{eq:twistdephasing}
\end{equation}
A finite tooth width therefore biases late-time observables. Narrowing
the teeth suppresses this bias but costs squeezing.

\subsection{Two-qumode circle and radial sectors}
\label{sec:circleencoding}

The second encoding we consider realizes the circle inside the plane of two
qumodes, with canonical pairs
$[\hat{q}_\mu,\hat{p}_\eta]=i\delta_{\mu\eta}\one$, with $\mu,\eta = 0,1$. The logic parallels
the GKP construction. The plane contains the circle as the constraint
surface $\hat{q}_0^2+\hat{q}_1^2=R^2\one$ of radius $R$ and we only
work with states concentrated on that
surface~\cite{Ale:2025sxz}. The angular momentum and the angular
phase are represented by
\begin{equation}
 \hat{J}=\hat{q}_0\hat{p}_1-\hat{q}_1\hat{p}_0,\qquad
 \hat{U}_R=\frac{\hat{q}_0+i\hat{q}_1}{R}.
 \label{eq:ringoperators}
\end{equation}
The exact unitary phase operator on the whole plane, $(\hat{q}_0+i\hat{q}_1)/\sqrt{\hat{q}_0^2+\hat{q}_1^2}$, requires reciprocals and square roots of quadratures, which are not available gates. We use $\hat{U}_R$ and absorb the constant $1/R$ into the Hamiltonian parameters through $\hat{L}\to\hat{J}$ and $\cos\hat{\theta}\to\hat{q}_0/R$. The commutator
$[\hat{J},\hat{U}_R]=\hat{U}_R$ holds on the whole plane, while
$\hat{U}_R^\dagger\hat{U}_R=(\hat{q}_0^2 + \hat{q}_1^2)/R^2 \overset{!}{=}\one$ holds only on the constraint surface,
where the substitutions reproduce the rotor Hamiltonian.

With $\hat{a}_\pm=(\hat{a}_0\mp i\hat{a}_1)/\sqrt2$, one obtains
\begin{align}
 \hat{J}&=i(\hat{a}_1^\dagger\hat{a}_0-\hat{a}_0^\dagger\hat{a}_1)
 =\hat{n}_+-\hat{n}_-,\nonumber\\
 &\text{with}\quad\hat{n}_{\pm}\equiv \hat{a}^{\dagger}_\pm\hat{a}_\pm.
 \label{eq:schwinger}
\end{align}
Integer angular momentum is therefore built into the planar
representation. The operator inequality $|\hat{J}|\le\hat{n}_++\hat{n}_-$ gives a
minimum occupation for each represented angular-momentum sector.

The linear substitution needs one refinement because if we naively insert it into
Eq.~\eqref{eq:hsV}, it produces terms linear in $\hat{q}_0$, so the
ground-diabat well, whose cosine enters with the positive coefficient
$W_0/2$, becomes unbounded from below on the plane. As in
Ref.~\cite{Ale:2025sxz}, we complete it to a form quadratic in the
quadratures,
\begin{equation}
 1-\cos\hat{\theta}\ \longmapsto\
 \frac{(\hat{q}_0-R\one)^2+\hat{q}_1^2}{2R^2},
 \label{eq:quadraticcosine}
\end{equation}
applied to the ground diabat. The excited diabat carries the cosine with
the opposite sign, so we write it about its own minimum at trans,
$-\frac{W_1}{2}(1-\cos\theta)=-W_1+\frac{W_1}{2}(1+\cos\theta)$, and
complete $1+\cos\hat{\theta}\mapsto[(\hat{q}_0+R\one)^2+\hat{q}_1^2]/(2R^2)$.
Both right-hand sides are nonnegative on the whole plane, so both diabats
are bounded from below, and on the constraint surface they agree exactly
with the linear substitution.

The exact structure is again a decomposition into conserved sectors since the Hamiltonian conserves the radius.
In the coordinate representation, write $q_0=R\rho\cos\theta$ and
$q_1=R\rho\sin\theta$. The pair $(R,\rho)$ mirrors the pair $(s,\nu)$
of the GKP encoding. The fixed radius $R$, like the scale $s$, is a
chosen parameter of the encoding: it never appears as an operator and
it sets physical costs such as the minimum occupation. The
dimensionless $\rho$ measures the radius in units of $R$, so the
constraint surface is $\rho=1$ and $\rho$ serves as the conserved fiber
label, with the offset $\rho-1$ playing the part of $\nu$. No radial
kinetic energy is added to the molecular Hamiltonian, so the radius is
conserved. Writing
$\hat{H}_x=\omega(\hat{p}_x^2+\hat{x}^2)/2$,
\begin{align}
 L^2(\mathbb R^2)&\simeq
 \int_0^\infty{}^{\oplus}\rho\,d\rho\,
 L^2_\rho(\mathbb S^1),\nonumber\\
 \hat{H}_\rho&=\frac{G}{2}\hat{J}^2+\hat{H}_x\nonumber\\
 &\quad+\hat{V}\Big(1\mp\cos\hat{\theta}\to
 \frac{1+\rho^2}{2}\mp\rho\cos\hat{\theta},\,\hat{x}\Big),
 \label{eq:radialsector}
\end{align}
where $\hat{J}$ acts as $-i\partial_\theta$ on each fiber and the
potential rule is Eq.~\eqref{eq:quadraticcosine} and its mirror evaluated
at radius $\rho$, with the upper sign in the ground diabat and the lower
sign in the excited diabat. Each radius $\rho$ is a fiber carrying a rotor Hamiltonian. Compared with Eq.~\eqref{eq:blochsector}, the kinetic term is now the same in every fiber and the potential varies with $\rho$. Because the constant $(1+\rho^2)/2$ enters the two diabats with the
coefficients $W_0/2$ and $W_1/2$, it shifts their energies differently, so the
electronic gap and the conical intersections move at first order in
$\rho-1$. As in the GKP case, the states of a single
fiber are non-normalizable with respect to the radial measure, each
radial label is conserved independently and a normalizable state
carries a conserved distribution over $\rho$ concentrated near the
physical fiber.

For the finite-energy preparation we replace the exact constraint by a
narrow annulus. Its radial amplitude is
\begin{equation}
 f_A(\rho)=\mathcal N_A
 \exp[-A^2(\rho^2-1)^2],
 \label{eq:ringstate}
\end{equation}
normalized with the measure $\rho\,d\rho$. This is the action of the
non-Gaussian filter $e^{-A^2(\hat{\rho}^2-\one)^2}$ on a smooth input
state, with $\hat{\rho}^2=(\hat{q}_0^2+\hat{q}_1^2)/R^2$. For $A\gg1$
the annulus has fractional radial width $\sigma_\rho\simeq1/(4A)$ and
physical width $\sigma_r\simeq R/(4A)$, which play the role of the
tooth width $\sigma_\nu$ of the GKP encoding.

The constraint filter has a direct ancilla construction. Let
$\hat{C}=\hat{q}_0^2+\hat{q}_1^2-R^2\one$, prepare an ancilla with momentum variance
$e^{-2r_a}/2$ and apply
\begin{equation}
 \hat{V}_a=e^{-i\gamma\hat{C}\hat{q}_a}.
 \label{eq:constraintgate}
\end{equation}
Conditioning on the central homodyne outcome gives the Kraus density
\begin{equation}
 \langle p_a=0|\hat{V}_a|\psi_a\rangle
 \ \propto\
 e^{-\gamma^2e^{2r_a}\hat{C}^2/2},\qquad
 \gamma e^{r_a}=\frac{\sqrt2 A}{R^2}.
 \label{eq:constraintfilter}
\end{equation}
For details of the circuit that builds this filter, see Ref.~\cite{Ale:2025sxz}.
Interaction strength and ancilla squeezing therefore enter through one
known combination, which sets the annular width. A physical acceptance
window has finite probability and slightly broadens the annulus.
Applied to an input carrying the molecular state
$\psi_{\rm mol}(\theta,x)$, the filter produces
Eq.~\eqref{eq:ringstate}.

For either encoding, the finite-width bias is an average of fiber trajectories over the conserved distribution $w(\nu)$ or $w(\rho)$. Conditioning the readout on the fiber label removes the bias at the cost of discarded shots. Sections~\ref{sec:collective} and~\ref{sec:counterterms} give two ways to reduce the cost of a narrow distribution, which can be combined.

\subsection{Collective encodings across multiple qumodes}
\label{sec:collective}

The first freedom is to distribute an encoded qumode over several
physical modes. For $k$ canonical pairs, define
\begin{equation}
 \hat{X}=\frac1{\sqrt k}\sum_{j=1}^k\hat{q}_j,\qquad
 \hat{P}=\frac1{\sqrt k}\sum_{j=1}^k\hat{p}_j,
 \qquad [\hat{X},\hat{P}]=i\one.
 \label{eq:collective}
\end{equation}
The unwrapped GKP coordinate is $\hat{\theta}_{\mathbb R}=s\hat{X}$,
with $\hat{L}_{\mathbb R}=\hat{P}/s$, or
\begin{equation}
 \hat{\theta}_{\mathbb R}=\frac{s}{\sqrt k}\sum_j\hat{q}_j,\qquad
 \hat{L}_{\mathbb R}=\frac1{s\sqrt k}\sum_j\hat{p}_j.
 \label{eq:collectivecanonical}
\end{equation}
An orthogonal passive transformation completes $(\hat{X},\hat{P})$
with $k-1$ difference modes~\cite{Reck:1994ujl}. Preparing these modes in
vacuum gives one collective encoded state with no additional occupied
degrees of freedom. Spreading a mode uniformly over many vacuum modes in
this way bounds the occupation that each mode must
represent~\cite{Oh:2026ljm}. Hamiltonians that depend only on $\hat{X}$ and $\hat{P}$
preserve this sector.

For the circle, we distribute each Cartesian pair in the same way
\begin{equation}
 \hat{q}_\mu=\frac1{\sqrt k}\sum_j\hat{q}_{\mu j},\qquad
 \hat{p}_\mu=\frac1{\sqrt k}\sum_j\hat{p}_{\mu j},\quad \mu=0,1.
 \label{eq:ringcollective}
\end{equation}
The constraint acts on $\hat{q}_0^2+\hat{q}_1^2$ and there are $2k$ physical
torsion modes. One can equivalently distribute the two circular modes,
provided the state and all operators are transformed consistently.
The construction adds empty difference modes and changes the mode
basis. It preserves both the molecular dynamics and the total number
distribution, while reducing the occupation required in each output
mode. The quantitative tradeoff and its gate implementation are derived
in Sec.~\ref{sec:resources}.

\subsection{Hamiltonian counterterms}
\label{sec:counterterms}

The second freedom lies in the Hamiltonian itself. The molecule fixes
the encoded Hamiltonian only on the physical fiber. Away from it we may
add terms, which we call counterterms. They must commute with the fiber
label, $\hat{S}$ of Eq.~\eqref{eq:gkpstabilizer} for the GKP encoding
and $\hat{\rho}^2$ for the circle, so that the label stays conserved and
the state still splits into independent fibers. They must also leave
the physical fiber unchanged, use the available gates and keep the
Hamiltonian bounded from below. Chosen to cancel the leading dependence
on the fiber label, they let a broader fiber distribution reach the
same accuracy. The encoding then needs less squeezing and fewer
photons.

For the GKP encoding the fiber label enters only through the kinetic
shift of Eq.~\eqref{eq:blochsector}. Replacing the encoded momentum in
the kinetic energy $G\hat{L}_{\mathbb R}^2/2$ by
\begin{equation}
 f_3(\hat{L}_{\mathbb R})=\hat{L}_{\mathbb R}
 -\frac{\sin(2\pi\hat{L}_{\mathbb R})}{2\pi}
 \label{eq:momentumflatten}
\end{equation}
removes this dependence at leading order. At momentum $\ell+\nu$,
$f_3=\ell+(2\pi^2/3)\nu^3+O(\nu^5)$, so $Gf_3(\hat{L}_{\mathbb R})^2/2$
reproduces the kinetic energy for every integer $\ell$ and the
coherence loss of Eq.~\eqref{eq:twistdephasing} scales with
$\langle\nu^6\rangle$ instead of $\sigma_\nu^2$. This kinetic phase is
not Gaussian and can be realized with the trigonometric gate synthesis
of Refs.~\cite{Rainaldi:2025ymn,Chalermpusitarak:2025cod}.
More generally, the trigonometric polynomial
\begin{align}
 &f_{2k+1}(\hat{L}_{\mathbb R})\nonumber\\
 &\quad=\hat{L}_{\mathbb R}+\frac{1}{\pi\binom{2k}{k}}\sum_{j=1}^{k}
 \frac{(-1)^j}{j}\binom{2k}{k-j}\sin(2\pi j\hat{L}_{\mathbb R}),
 \label{eq:momentumflattenk}
\end{align}
reduces to Eq.~\eqref{eq:momentumflatten} for $k=1$. Its derivative
$f_{2k+1}'(\hat{L}_{\mathbb R})=4^k\sin^{2k}(\pi \hat{L}_{\mathbb R})/\binom{2k}{k}$ is nonnegative and
vanishes to order $2k$ at every integer, so $f_{2k+1}$ leaves every
integer fixed and $f_{2k+1}(\ell+\nu)=\ell+O(\nu^{2k+1})$. It cancels
the fiber dependence through order $\nu^{2k}$ at the cost of $k$
trigonometric harmonics. As $k$ grows, $f_{2k+1}$ approaches the
staircase that rounds $L_{\mathbb R}$ to the nearest integer.

For the circle embedding, the exact fiber-blind cosine would divide $\hat{q}_0$ by the radial
coordinate $\hat{\rho}$. Since $\hat{\rho}^2=(\hat{q}_0^2+\hat{q}_1^2)/R^2$ is a
polynomial, we expand $1/\hat{\rho}=(\hat{\rho}^2)^{-1/2}$ in powers of
$\hat{\rho}^2-\one$ and truncate at order $k$, such that
\begin{equation}
 \hat{C}_{2k+1}=\hat{q}_0\sum_{n=0}^{k}\frac{(-1)^n}{4^n}\binom{2n}{n}
 \left(\hat{\rho}^2-\one\right)^n
 \label{eq:counterterm}
\end{equation}
is a polynomial of degree $2k+1$ in the quadratures. On the fiber of
radius $\rho=1+\delta$, $\hat{C}_{2k+1}/R=g_{2k+1}(\rho)\cos\hat{\theta}$
with
\begin{equation}
    g_{2k+1}(1+\delta)=1+\frac{(-1)^k}{2^{k+1}}\binom{2k+2}{k+1}\delta^{k+1}
+O(\delta^{k+2}),
\end{equation}
so the cosine strength is fiber blind through order
$\delta^k$. The lowest order $k=1$ gives
$\hat{C}_3=\frac32\hat{q}_0-\frac{1}{2R^2}\hat{q}_0
(\hat{q}_0^2+\hat{q}_1^2)$ with
$g_3(1+\delta)=1-\frac32\delta^2-\frac12\delta^3$. In an experiment the
coefficients can also be calibrated by tuning them until the
fiber-resolved observable is stationary across the populated fibers.

The fiber constant is flattened in the same way, which also restores
boundedness, since the odd tail of $\hat{C}_{2k+1}$ alone would be
unbounded. The completed potential is replaced by
\begin{equation}
 1-\cos\hat{\theta}\ \longrightarrow\
 \one+\eta\,(\hat{\rho}^2-\one)^{2m}-\frac{\hat{C}_{2k+1}}{R},
 \label{eq:boundedflatten}
\end{equation}
with $m=\lceil(k+1)/2\rceil$ and $\eta>0$. On every fiber this equals
$1+\eta(\rho^2-1)^{2m}-g_{2k+1}(\rho)\cos\hat{\theta}$, which matches
$1-\cos\hat{\theta}$ on the physical fiber through order $\delta^k$.
Since $4m>2k+1$, the radial penalty dominates the counterterm at large
radius, so the ground-diabat well is bounded from below. Its strength
$\eta$ is a free parameter that leaves the physical fiber unchanged and
only sets the radius beyond which the potential is bounded. Because the
radius is conserved, the penalty does not change the radial
distribution of the state. In the fiber propagations of
Appendix~\ref{app:numerical} we set $k=1$ and $\eta=1/4$. For the excited
diabat we replace $1+\cos\hat{\theta}$ by
$\one+\eta(\hat{\rho}^2-\one)^{2m}+\hat{C}_{2k+1}/R$, the same operator with
the sign of $\hat{C}_{2k+1}$ reversed, which is bounded for the same reason.
Like the redistribution over modes, the counterterms leave the
molecular dynamics on the physical fiber unchanged. Their effect on
the squeezing and the occupation is quantified in
Sec.~\ref{sec:resource:occupation} and their gate implementation in
Sec.~\ref{sec:resource:gates}.

%% file: sections/results.tex
\section{Hahn-Stock dynamics and finite-resource benchmarks}
\label{sec:results}

\subsection{Local approximations and compact dynamics}
\label{sec:hsresults}

Before working with the bosonic encodings of the previous sections, we benchmark the continuum molecular dynamics across four Hamiltonian descriptions to isolate two distinct physical effects: the breakdown of local potential expansions during large-amplitude motion and the dynamical signatures of compact torsional topology. We then show that adding Fock levels to a direct qumode encoding cannot recover the compact dynamics, while the compact encodings of Sec.~\ref{sec:encodings} do.

\begin{figure*}[t]
 \centering
 \includegraphics[width=\textwidth]{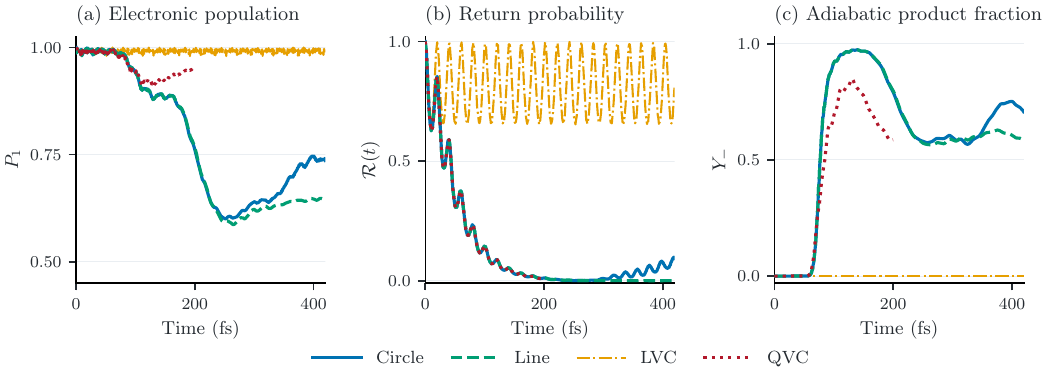}
 \caption{Hahn-Stock dynamics for local and global descriptions of the torsion. (a) Electronic population $P_1$. (b) Return probability $\mathcal R(t)$ of Eq.~\eqref{eq:return}. (c) Conditional lower-adiabatic trans probability $Y_-=P({\rm trans},S_0)/P(S_0)$, with the adiabatic projector of each model's own potential. All curves start from the same packet in the cis basin and the line readout folds probabilities over cells. The LVC and QVC controls use Eqs.~\eqref{eq:lvc} and \eqref{eq:qvc}, with QVC shown through 200 fs (Sec.~\ref{sec:hsresults}). The trans readout assigns half weight to the discrete basin-boundary nodes.}
 \label{fig:hsdynamics}
\end{figure*}

In Fig.~\ref{fig:hsdynamics} we compare the time evolution of a wavepacket initialized in the cis basin under the linear vibronic coupling (LVC) model, the quadratic vibronic coupling (QVC) model, the full trigonometric potential on the noncompact line $\R$ and the physical rotor Hamiltonian on the compact circle $\Sone$. The torsional coordinate in the LVC model decouples from the electronic and vibrational degrees of freedom, as Eq.~\eqref{eq:lvc} implies. Consequently, the wavepacket exhibits weak electronic-vibrational oscillations while remaining near the cis geometry.

The QVC model restores the negative torsional curvature on the excited diabat, capturing the initial acceleration away from the cis region and the first passage through the crossing region. At later times, its inverted excited-state parabola drives the wavepacket outward without bound. QVC then departs from the full-potential calculations after about 100 fs. Because the QVC Hamiltonian is unbounded from below along the torsion, the escaping packet keeps gaining kinetic energy and spreads exponentially in time, so the number of grid points or Fock levels needed to follow it grows exponentially with the propagation time. We therefore stop QVC at 200 fs. Longer times would add no information, since by then QVC has already departed from the full potential. We propagate Eq.~\eqref{eq:qvc} on a wide position grid, so the deviation of QVC from the compact calculation reflects the model error alone.

Comparing the full trigonometric potential on $\R$ and $\Sone$ isolates the role of the global topology in the time evolution. Because the initial wavepacket is tightly localized ($\sigma_{\theta,0}\ll\pi$) and both manifolds share identical local forces, the line reference accurately tracks the physical circle dynamics at early times. The two trajectories diverge once counter-propagating branches interfere: the circle retains the cross terms between amplitudes in different angular cells, while the folded line does not. Eq.~\eqref{eq:windinginterference} makes this explicit at the wavefunction level. A packet launched near the top of the excited torsional potential spreads in both angular directions. On the circle the two branches meet after traversing different paths and interfere, while on the line their copies occupy distinct cells. With all potential terms held fixed, the electronic population on the line stays within $0.01$ of the circle until about 256 fs before diverging, reaching a difference of $0.10$ near 400 fs, ten times the tolerance. Similarly, the conditional adiabatic product fraction $Y_-$ differs by up to $0.15$.
The return probability shows the topology most directly. On the line the copies of the packet move apart in different cells and $\mathcal R(t)$ decays to zero after about 250 fs. On the circle the two branches come back to the Franck-Condon region and $\mathcal R(t)$ revives to about $0.09$ by 420 fs. The line reference is numerically converged and verified across two independent implementations to far below these topological discrepancies (Appendix~\ref{app:numerical}).

\subsection{Cutoff effects in direct qumode encodings}
\label{sec:directcutoff}

The simplest way to place the torsion on a qumode is to identify the angle with a quadrature and keep the full cosine potential. We call this a direct encoding. The quadrature is not periodic, so a direct encoding simulates the line and not the circle. A finite register adds a cutoff error on top of it. Here and below, $N$ denotes the number of Fock states $\{\ket{0},\dots,\ket{N-1}\}$ retained per physical qumode. These are distinct
from the rotor levels $\ket{\ell}$ of the encoded torsion.

\begin{figure}[t]
 \centering
 \includegraphics[width=\columnwidth]{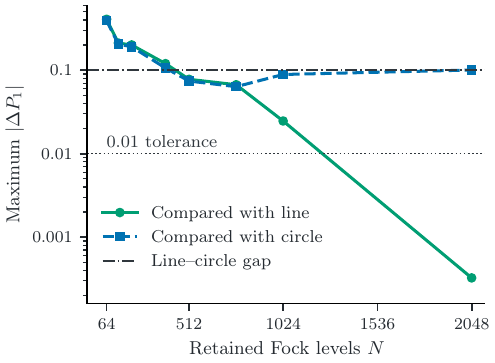}
 \caption{ Fock cutoff of a direct qumode encoding of the torsion in Hahn-Stock dynamics. We show the largest error in $P_1$ over $420$ fs when the noncompact trigonometric Hamiltonian is truncated to $N$ Fock levels at the qumode scale $s=0.32$ (Appendix~\ref{app:numerical}). The green line compares each register with the untruncated noncompact theory and shows how the noncompact theory converges onto itself as $N$ grows. The dashed blue line compares each register with the compact rotor and tends to the dash-dotted line, the largest gap between the noncompact and compact theories. The dotted line marks the $0.01$
 tolerance.}

 \label{fig:truncationtopology}
\end{figure}

In Fig.~\ref{fig:truncationtopology} we compare each finite register with the line and circle references through the maximum population error $\epsilon_{P_1}$. The line reference is the limit $N\to\infty$ of the register, so the error against it measures the truncation alone. This error decreases steadily with the cutoff and reaches $3\times10^{-4}$ at $N=2047$. The error against the circle instead approaches the largest difference between the line and circle references, $0.10$, shown by the dash-dotted line, and reaches it at $N=2047$. At intermediate cutoffs it falls below this value because the truncation error and the difference between line and circle have opposite signs at late times and partly cancel. More Fock levels therefore only converge the simulation to the unwrapped line. Recovering the physical photoisomerization dynamics requires encoding the compact $\Sone$ topology directly.

\subsection{Finite energy and Fock truncation}
\label{sec:finiteresults}

We now check that the two compact encodings reproduce the circle with finite resources. A physical implementation has a finite fiber width, set by the squeezing or the constraint strength, and a finite Fock cutoff per qumode. With a single register per encoded mode, at $r_{\rm eff}=2.1$ and qumode scale $s=0.6$ for the GKP comb and at $A=192$ and $R=20$ for the circle, both encodings reproduce $P_1$, $Q_-$, $Y_-$ and the return probability $\mathcal{R}$ within $0.01$ through $416.5$ fs. For the circle the error in $Y_-$ includes a bound for the probability missed by the readout (Appendix~\ref{app:numerical}). These single registers need large cutoffs and strong squeezing. In Sec.~\ref{sec:resources} we show how spreading the encoding overseveral qumodes and adding counterterms bring them within a budget of 13 dB of squeezing and about $100$ Fock levels per mode.

%% file: sections/psb3.tex
\section{PSB3: a three-coordinate molecular application}
\label{sec:psb3}

To demonstrate how our hybrid framework scales to multi-coordinate reaction pathways, we apply it to the protonated Schiff-base model PSB3 of the retinal chromophore \cite{marsili2019, marsili2020,aarabi2025}. It shares with Hahn-Stock the same three types of degrees of freedom: electronic states (qubits), noncompact vibrational coordinates (oscillators) and a compact torsion (a rotor). The encodings of Sec.~\ref{sec:encodings} therefore carry over directly. However, PSB3 introduces three complications: the torsion couples simultaneously to two distinct vibrational modes, the potential depends on the torsion through more than one cosine term and the global minimum of the lower adiabatic surface $S_0$ lies at the trans conformer rather than the cis reactant.

\subsection{Coordinates and full molecular Hamiltonian}
\label{sec:psb3:model}

In addition to the central torsion $\hat\theta$, the model includes a bond-length alternation (BLA) coordinate $\hat r$ and a hydrogen-out-of-plane (HOOP) coordinate $\hat\phi$. The electronic coupling and the diagonal potentials depend on the combination $\hat\tau=\hat\theta-\hat\phi/2$, so torsional motion changes the forces acting on the other coordinates. We use the refined two-state potential of Refs.~\cite{marsili2019,marsili2020,quantummodellib}. The compact molecular
Hamiltonian is
\begin{align}
 \hat H_{\rm PSB3}
 &=\hat T\one_2+\hat V_{\rm PSB3}(\hat r,\hat\theta,\hat\phi),
 \label{eq:psb3hamiltonian}\\
 \hat T&=\frac12\left(G_{rr}\hat p_r^2+
 G_{\theta\theta}\hat L^2+G_{\phi\phi}\hat p_\phi^2\right)
 \nonumber\\
 &=-\frac12\left(G_{rr}\partial_r^2+
 G_{\theta\theta}\partial_\theta^2+
 G_{\phi\phi}\partial_\phi^2\right),
 \label{eq:psb3kinetic}
\end{align}
where $\hat L=-\ii\partial_\theta$ acts on periodic wavefunctions while $\hat p_r=-\ii\partial_r$ and $\hat p_\phi=-\ii\partial_\phi$ act on the two noncompact coordinates. The coordinate-space matrix of $\hat V_{\rm PSB3}$ is denoted by $V(r,\theta,\phi)$ below.
We use the diagonal kinetic approximation of Ref.~\cite{aarabi2025}, with coefficients listed in Table~\ref{tab:psb3parameters} of Appendix~\ref{app:psb3}. Reference~\cite{marsili2020} also considers a torsion-HOOP kinetic cross term. Here the diagonal metric and the two noncompact coordinate ranges are held fixed across the potential and topology comparisons.

To express the potential operator $\hat{V}_{\rm PSB3}$, we denote the BLA coordinate in angstroms by $r_{\rm A}$, while everywhere else $r$ is in bohr. We define the Morse and quadratic radial functions
\begin{align}
 M_P(r_{\rm A})&=D\left[1-\e^{-d_1(r_{\rm A}-r_{\rm TS})}\right]^2,
 \nonumber\\
 M_M(r_{\rm A})&=D\left[1-\e^{-d_4(r_{\rm A}-r_{\rm min})}\right]^2,
 \nonumber\\
 C(r_{\rm A})&=c_1r_{\rm A}^2+c_2r_{\rm A}+c_3.
 \label{eq:psb3radialfunctions}
\end{align}
In the diabatic basis $\{\ket{0},\ket{1}\}$, we write the matrix elements of $\hat V_{\rm PSB3}$ directly in Fourier components of
$\hat\tau$,
\begin{align}
 \hat V_{00}&=A_0(\hat r,\hat\phi)+\frac{d_3}{2}\cos\hat\tau
 +A_2(\hat r)\cos2\hat\tau,
 \nonumber\\
 \hat V_{11}&=F_0(\hat r,\hat\phi)+F_2(\hat r)\cos2\hat\tau,\nonumber\\
 \hat V_{01}&=k_1\sin2\hat\tau,
 \label{eq:psb3fourier}
\end{align}
with
\begin{align}
 A_0&=\frac{M_P+d_2+M_M+d_3}{2}+h_1\sin^2(\phi/4),\nonumber\\
 A_2&=\frac{M_M-M_P-d_2}{2},\nonumber\\
 F_0&=\left(1+\frac{c_5}{2}\right)C+\frac{c_4}{2}
       +h_1\sin^2(\phi/4),\nonumber\\
 F_2&=\frac{c_4-c_5C}{2},
 \label{eq:psb3coefficients}
\end{align}
with the parameters of Table~\ref{tab:psb3parameters}. In this representation, the non-Gaussian trigonometric interactions act on linear combinations of the torsional and HOOP coordinates, such as $\cos(2\hat\theta-\hat\phi)$ and $\cos(\hat\theta-\hat\phi/2)$. While the Morse functions are exponentials of $\hat r$, which no native gate can implement, the wavepacket explores a BLA interval short compared to the Morse decay lengths $d_{1,4}^{-1}$. For gate synthesis, these radial factors can therefore be represented by a low-order power series in $\hat r$. Alternatively, one can use a trigonometric expansion with the multimode Fourier synthesis of Ref.~\cite{Chalermpusitarak:2025cod}. In all numerical calculations in this section we retain the exact Morse potentials.

We show both adiabatic surfaces, $S_0$ and $S_1$, at $\phi=0$ over the full periodic domain $\theta\in[-\pi,\pi]$ in Fig.~\ref{fig:psb3}(a). Solving $V_{01}(r,\theta,0)=0$ and $V_{00}(r,\theta,0)=V_{11}(r,\theta,0)$ locates the two conical intersections at $\theta=\pm\pi/2$ and $r\simeq0.05139$ bohr, which enable nonadiabatic branching between the cis ($\theta=0$) and trans ($\theta=\pm\pi$) basins.

\subsection{Cis preparation and comparison Hamiltonians}
\label{sec:psb3:preparation}

Unlike in the Hahn-Stock model, the global minimum of the PSB3 ground adiabatic surface $S_0$ lies in the trans basin, while the cis reactant occupies a higher-energy metastable well. Both minima lie at $r_{\rm A}=r_{\rm min}$ and $\phi=0$. The cis minimum lies $d_3=0.165$ eV above the trans one. Along the minimum-energy torsional path, the two wells are separated by a barrier at $\theta=\pi/2$ that lies $2$ eV above the cis minimum and keeps the cis conformer metastable. 

However, in the eye, interactions with the surrounding protein keep retinal, of which PSB3 is a reduced model, in the cis form until a photon arrives. The photon then rotates the molecule to trans, and this motion starts the chain of events that produces a nerve signal~\cite{schoenlein1991,polli2010}. The cis well is therefore the physically relevant starting point.

Since the ground state is localized in the trans region, we isolate the cis-localized state that occurs in retinal by adding a confining wall

\begin{equation}
 W_{\rm cis}(\theta)=0.50\min\left[
 \left(\frac{(|\theta|-\theta_w)_+}{\pi-\theta_w}\right)^4,1\right]
 \quad\text{Hartree},
 \label{eq:psb3ciswall}
\end{equation}
with $\theta_w=0.48\pi$ and $(z)_+=\max(z,0)$. This function vanishes throughout the cis well, raises the trans basin out of reach and defines the multiplication operator $\hat W_{\rm cis}$.

If $\ket{u_-(q)}$ and $\ket{u_+(q)}$ are the local electronic eigenvectors of $V(q)$, a sudden vertical photoexcitation transfers the ground packet from the lower to the upper adiabatic surface at every geometry and prepares the initial wavefunction
\begin{align}
 \Psi_{\rm FC}(q)&=\mathcal N\hat D_{\rm FC}(q)\Psi_g(q),
 \nonumber\\
 \hat D_{\rm FC}(q)&=\ket{u_+(q)}\bra{u_-(q)},
 \qquad q=(r,\theta,\phi).
 \label{eq:psb3preparation}
\end{align}

The wall is then removed and $\Psi_{\rm FC}$ is propagated under $\hat H_{\rm PSB3}$. Preparing this state on hardware is a nontrivial task which lies outside the scope of this work. Adaptive variational methods acting natively on qumodes, such as CV-ADAPT-VQE~\cite{Athanasakos:2026upp}, suggest a possible route toward preparing the initial ground packet.

We benchmark the exact compact dynamics on $\Sone$ against two approximations sharing the same initial state $\Psi_{\rm FC}$ and diagonal kinetic operator $\hat{T}$:
\begin{enumerate}
 \renewcommand{\labelenumi}{(\roman{enumi})}
 \item The quadratic vibronic coupling (QVC) model, the PSB3 counterpart of Eq.~\eqref{eq:qvc}. We Taylor-expand the potential matrix $V(r,\theta,\phi)$ to second order in $\theta$ around $\theta=0$ while keeping the full nonpolynomial dependence on $r$ and $\phi$ intact:
\begin{align}
 V_{\rm QVC}(r,\theta,\phi)&=V(r,0,\phi)
 +\theta\left.\partial_\theta V\right|_{\theta=0}\nonumber\\
 &\quad+\frac{\theta^2}{2}
 \left.\partial_\theta^2V\right|_{\theta=0}.
 \label{eq:psb3local}
\end{align}
 This isolates the breakdown of local torsional approximations without introducing BLA or HOOP truncation errors.
 \item A noncompact trigonometric line. We unwrap $\theta$ onto $\R$ using a converged five-cell periodic box of length $10\pi$, folding the resulting adiabatic probability densities modulo $2\pi$ onto $[-\pi,\pi]$ (with half weight assigned to the basin-boundary nodes at $|\theta|=\pi/2$).
\end{enumerate}

\subsection{Dynamics and finite-resource results}
\label{sec:psb3:results}

\begin{figure*}[t]
 \centering
 \includegraphics[width=\textwidth]{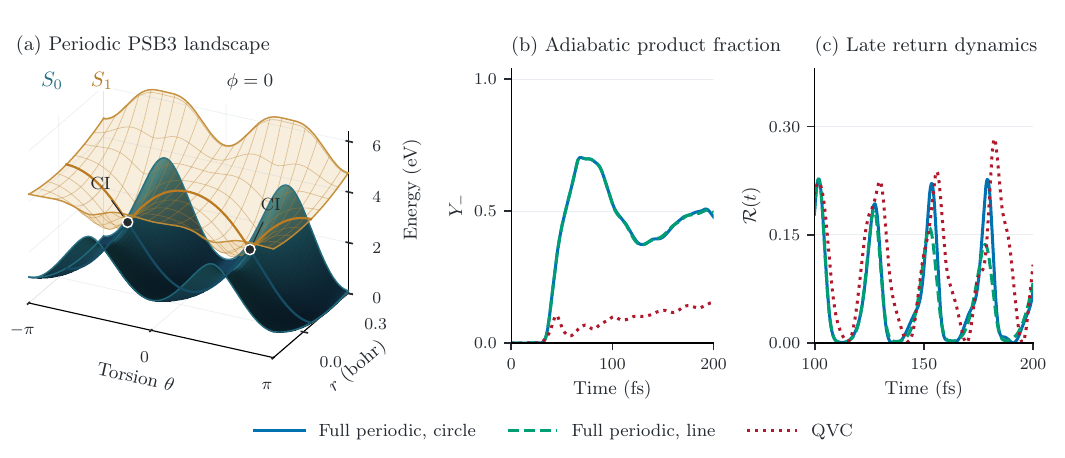}
 \caption{Molecular geometry and dynamics of the three-coordinate PSB3
 model. (a) Lower and upper adiabatic surfaces at $\phi=0$, evaluated
 over the full period $-\pi\le\theta\le\pi$ and
 $-0.20\le r\le0.30$ bohr. The markers labeled CI are the two conical
 intersections.
 (b) Conditional lower-adiabatic trans probability
 $Y_-=P({\rm trans},S_0)/P(S_0)$.
 (c) Late return probability to the initial molecular state.
 The physical circle dynamics are compared with an unwrapped
 five-cell line reference using the same periodic potential. The QVC
 model expands the torsional
 trigonometric factors through second order and preserves the HOOP
 dependence.}
 \label{fig:psb3}
\end{figure*}

In Figs.~\ref{fig:psb3}(b) and \ref{fig:psb3}(c) we compare these three descriptions for the same cis-localized preparation over a time window of $200$ fs. Just as in the Hahn-Stock model, the QVC model $V_{\rm QVC}$ misses the reaction, with a product fraction that differs from the exact conditional product fraction $Y_-(t)$ by up to $0.64$. Over the $200$~fs window, the noncompact line $\R$ and the compact circle $\Sone$ differ by at most $0.016$ in $Y_-(t)$ and $0.019$ in the adiabatic $S_1$ population.

Compactness gives a larger signal in the late return probability $\mathcal{R}(t)$, where the line and circle differ by $0.10$ near 180 fs. The return and product observables thus probe different aspects of the motion. Conditional-state overlaps between winding paths (Eq.~\eqref{eq:windinginterference}) can resolve how the additional coordinates shape this difference.

Finally, we verify that our compact bosonic encodings accurately capture the three-mode PSB3 dynamics under finite resources. The finite-energy GKP comb at $r_{\rm eff}=1.5$ holds the product fraction within $0.003$ of the compact dynamics through 200 fs. To test Fock-space truncation in the presence of all three coupled modes, we retain $3072$ Fock levels in the torsional register and propagate through the first $70$~fs. We read out after a normalized projection onto the 256 GKP teeth, which discards at most $0.0074$ of the probability. The encoded trajectory then tracks the exact $\Sone$ reference to within $0.004$ in the adiabatic $S_1$ population and $0.008$ in $Y_-(t)$. This comparison of the product fraction uses a strict basin boundary. Extending the truncated dynamics into the late-time recurrence regime is the natural next resource test for this model. We leave this to future work. 

%% file: sections/resources.tex
\section{Hardware resources and experimental prospects}
\label{sec:resources}

In this section, we determine the qumode and gate resources required to implement the compact encodings of Sec.~\ref{sec:encodings} on hybrid quantum hardware. While the single-register benchmarks of Sec.~\ref{sec:results} require large Fock cutoffs and high squeezing, distributing the encoding across multiple qumodes and incorporating Hamiltonian counterterms or postselection lowers the required squeezing and per-mode occupation. The numbers below are estimates for stated assumptions and initial conditions, and we use them to explain the logic of the resource count rather than to fix a hardware target.

\subsection{Qumode frames, localization and occupation}
\label{sec:resource:frame}

For the GKP encoding with the unwrapped qumode angle $\hat{\theta}_{\mathbb R}=s\hat{q}$ and $\hat{L}_{\mathbb R}=\hat{p}/s$, the occupation of a fixed encoded state is exactly
\begin{align}
 \langle\hat{n}\rangle_s(t)&=\frac12
 \left[\frac{A_\theta(t)}{s^2}+s^2B_\theta(t)-1\right],\nonumber\\
 A_\theta&=\langle\hat{\theta}_{\mathbb R}^2\rangle,\qquad
 B_\theta=\langle\hat{L}_{\mathbb R}^2\rangle.
 \label{eq:gkpnumber}
\end{align}
Balancing the two terms gives the optimal $s_{\rm opt}^4=A_\theta/B_\theta$ and the occupation $\sqrt{A_\theta B_\theta}-1/2$. Because a device uses one choice of $s$ for the whole trajectory, the register capacity is dictated by the largest occupation over the time window. Furthermore, an upper bound $r_{\rm phys}\le r_{\max}$ on the physical squeezing of the teeth imposes, through Eq.~\eqref{eq:physicalwidth},
\begin{equation}
 s\ge s_{\min}=\sqrt{2\pi}\,e^{r_{\rm eff}-r_{\max}}.
 \label{eq:squeezingconstraint}
\end{equation}
Less squeezing at fixed tooth width $\sigma_\nu$ of Eq.~\eqref{eq:width} therefore requires a larger $s$ and pushes the momentum teeth $p=s\ell$ farther from the origin. Once $s_{\min}$ exceeds $s_{\rm opt}$, this raises the boson occupation through the term $s^2B_\theta$ of Eq.~\eqref{eq:gkpnumber}.

For the two-qumode circle encoding at radius $R$ and physical radial width $\sigma_r\simeq R/(4A)$, we obtain the mean occupation number
\begin{equation}
 \langle\hat{n}_0+\hat{n}_1\rangle
 \simeq\frac{R^2}{2}+\frac1{8\sigma_r^2}
       +\frac{\langle\hat{J}^2\rangle}{2R^2}-1,
 \label{eq:ringestimate}
\end{equation}
where the three positive terms correspond to the radial displacement, the radial momentum and the angular momentum. A visually narrow annulus can therefore be expensive in the mean occupation number. At fixed radial width the first and third terms compete. Minimizing over $R^2$ gives the photon-optimal radius $R^2=\sqrt{\langle\hat{J}^2\rangle}$ and the occupation $\sqrt{\langle\hat{J}^2\rangle}+1/(8\sigma_r^2)-1$. The radius cannot always take this value, because the dynamics constrains the fractional width $\sigma_\rho=\sigma_r/R$, which sets the spread of the fiber label. A tolerance $\sigma_\rho\le\sigma_\rho^{\max}$ requires $R\ge\sigma_r/\sigma_\rho^{\max}$, so the radius is
\begin{equation}
 R=\max\left(\langle\hat{J}^2\rangle^{1/4},\
 \sigma_r/\sigma_\rho^{\max}\right).
 \label{eq:ringradius}
\end{equation}
Relaxing $\sigma_\rho^{\max}$, as the counterterms of Sec.~\ref{sec:counterterms} do, lets the circle reach its photon-optimal radius (Fig.~\ref{fig:multimode}).

\subsection{Trading qumode count for qumode occupation}
\label{sec:resource:occupation}

Even at the optimal frame a single qumode may hold hundreds of photons for the trajectories of Sec.~\ref{sec:results}. The first freedom of Sec.~\ref{sec:collective} is the number of physical modes we use. We can organize each encoded coordinate in a collective mode $\hat{a}_{\rm coll}=k^{-1/2}\sum_j\hat{a}_j$ of $k$ qumodes, with all orthogonal modes in vacuum. The wrapped encoding uses one such mode and the circle a pair, so the circle uses $2k$ physical modes. This is simply a passive mixing that conserves the total occupation. If the collective mode has number distribution $p_n$, each of its photons lands in a given physical mode with probability $1/k$, so that mode holds $m$ photons with probability
\begin{equation}
 q_k(m)=\sum_{n=m}^\infty p_n\binom{n}{m}
               k^{-m}(1-k^{-1})^{n-m},
 \label{eq:resource:binomial}
\end{equation}
also when the collective mode is entangled with the molecule. The same binomial splitting underlies multiplexed photon-number-resolving detection~\cite{Fitch:2003ppy,Achilles:2003vzu}. Its mean is $\bar n/k$ for a collective mean $\bar n$. Sharing a tail budget $\epsilon$ among the $k$ marginals, so that each mode misses at most $\epsilon/k$, the Fock levels needed per mode follow for a narrow collective distribution from the normal approximation
\begin{equation}
 N-1\simeq\frac{\bar n}{k}
 +z\sqrt{\frac{\bar n}{k}\left(1-\frac1k\right)},
 \label{eq:splitquantile}
\end{equation}
with $z$ the upper-tail quantile of $\epsilon/k$. The exact mixture of Eq.~\eqref{eq:resource:binomial} applies when the distribution is broad.

To illustrate the resources we require the electronic population $P_1$, the direct qubit readout, to stay within $0.03$ of the compact dynamics through 420 fs for both encodings. We take 13 dB of squeezing and 101 Fock levels per qumode as an example budget. We choose an odd number of Fock states so that the zero eigenvalue for both quadratures exists.

\begin{figure}[t]
\centering
\includegraphics[width=\columnwidth]{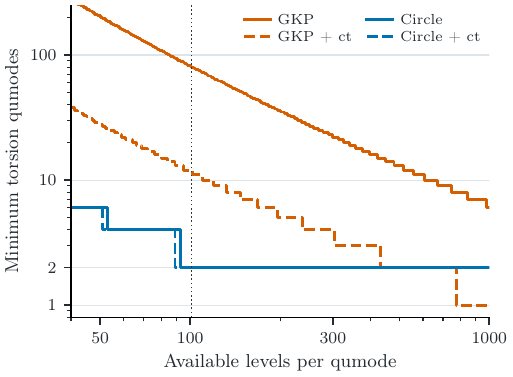}
\caption{Minimum torsion qumodes for a given number of Fock levels per qumode at a 13 dB squeezing cap, with fiber tolerances that hold $P_1$ within $0.03$ through 420 fs, without (solid) and with (dashed) the counterterms (ct) of Sec.~\ref{sec:counterterms}.
At 101 levels (dotted line) the GKP comb needs 80 qumodes for about 5300 photons and the kinetic counterterm reduces this to 12 qumodes for about 770 photons. The circle needs a radius $R=14.2$ without the counterterm and $R=8.45$ with it, and in both cases fits on its two native qumodes.}
\label{fig:multimode}
\end{figure}

At the single-register operating points of Sec.~\ref{sec:finiteresults} the mean occupation peaks at $837$ for the GKP comb along its 420 fs trajectory and at $406$ in total for the circle, $203$ in each circular mode. Under the 13 dB cap the occupation is instead set by the fiber tolerance each encoding must meet. Figure~\ref{fig:multimode} shows how many qumodes each encoding then needs for a given number of Fock levels per qumode, without and with its counterterm.

The GKP curves of Fig.~\ref{fig:multimode} follow from Eqs.~\eqref{eq:gkpnumber} and~\eqref{eq:squeezingconstraint} at the fiber tolerances of the comb without and with the kinetic counterterm (Appendix~\ref{app:numerical}). They use the moments of the propagated GKP trajectory of Sec.~\ref{sec:finiteresults} with one frame for the whole window and are split over modes by Eq.~\eqref{eq:splitquantile}. The circle curves are analytic displaced-squeezed distributions at a 13 dB cap, with the radii fixed by the fiber tolerances of Appendix~\ref{app:numerical} and the angular term of Eq.~\eqref{eq:ringestimate} at $\langle\hat{J}^2\rangle=\ell_E^2/2$, the energetic bound, which exceeds the largest $\langle\hat{L}^2\rangle\simeq2840$ of that trajectory.
Every plotted cutoff keeps the probability outside the truncated register below $10^{-3}$, shared equally among its modes. These curves are occupation estimates for the stated tolerance, squeezing cap, initial state and time window. They change with these assumptions. A truncated register conserves the fiber label only approximately, so we propagate the truncated two-mode circle register with its counterterm. At 101 levels per mode it holds $P_1$ within the $0.03$ target through 420 fs.

We can also always postselect the readout on the conserved fiber label. Keeping only the shots near the physical fiber removes most of the finite-width bias, so a broader fiber distribution, with less squeezing and fewer photons, reaches the same accuracy at the cost of repeated shots.

Redistribution leaves the tooth width, the radial resolution and the total occupation unchanged, so it replaces one demanding qumode with several less occupied ones at equal mean loss exposure. The gate constructions below act directly on the distributed coordinates and preserve this advantage during the nonlinear evolution.

\subsection{Gate decomposition of the hybrid time evolution}
\label{sec:resource:gates}

We now construct the time evolution explicitly for the Hahn-Stock Hamiltonian.
We use the hybrid gate set of Ref.~\cite{Araz:2024dcy}: qubit rotations ${\rm R}_i(\phi)=e^{i\phi\hat{\sigma}_i/2}$, the qumode rotation ${\rm R}(\phi)=e^{i\phi\hat{n}}$, the displacement ${\rm D}(z)=e^{z\hat{a}^\dagger-z^*\hat{a}}$, the beam splitter ${\rm BS}(z)=e^{z\hat{a}^\dagger\hat{b}-z^*\hat{a}\hat{b}^\dagger}$, the Kerr and cross-Kerr phases ${\rm K}(\phi)=e^{i\phi\hat{n}^2}$ and ${\rm CK}(\phi)=e^{i\phi\hat{n}_a\hat{n}_b}$ and the conditioned displacement
\begin{equation}
 {\rm CD}(z)=e^{\hat{\sigma}_z(z\hat{a}^\dagger-z^*\hat{a})},
 \label{eq:cdconvention}
\end{equation}
with $z\in\mathbb C$ and $\phi\in\mathbb R$. We add the quadratic phase ${\rm P}(\kappa)=e^{i\kappa\hat{q}^2/2}$, its conditional form ${\rm CP}(\kappa)=e^{i\kappa\hat{\sigma}_z\hat{q}^2/2}$~\cite{Weedbrook:2011wxo} and the conditional cosine gate ${\rm Cos}(\gamma)=e^{-i\gamma\cos(s\hat{q})\otimes\hat{\sigma}_z}$ of Refs.~\cite{Rainaldi:2025ymn,Chalermpusitarak:2025cod,Rainaldi:2026wpt}. Their pulse realizations depend on the platform \cite{Lloyd:1998jk,Eickbusch:2021uod,Rainaldi:2026wpt,McGarry:2026mkk}.

\begin{figure*}[t]
\centering
\small
\begin{minipage}[t]{0.47\textwidth}
\centering
(a) Wrapped encoding\\[4pt]
\resizebox{\linewidth}{!}{%
\begin{quantikz}[row sep=0.22cm,column sep=0.17cm,classical gap=0.02cm]
\lstick{ancilla $\ket{0}$} & & & & & & \ctrl{2} & & \\
\lstick{electronic $e$} & & \gate{{\rm R}_y(\beta)} & \ctrl{2} & \gate{{\rm R}_y(-\beta)} & \gate{{\rm R}_z} & & \ctrl{1} & \\
\setwiretype{b}\lstick{torsion $\theta$} & \gate{{\rm P}_p} & & & & & \gate{\rm Cos} & \gate{\rm Cos} & \\
\setwiretype{b}\lstick{vibration $x$} & \gate{\rm D} & & \gate{\rm CD} & & \gate{\rm R} & & &
\end{quantikz}}
\end{minipage}\hfill
\begin{minipage}[t]{0.47\textwidth}
\centering
(b) All qubits\\[4pt]
\resizebox{\linewidth}{!}{%
\begin{quantikz}[row sep=0.22cm,column sep=0.22cm]
\lstick{vibration $x$} & \qwbundle{m_x} && \gate{\hat B_x} & \gate[2]{e^{-i\tau\hat M}} & \gate{\hat B_x^\dagger} & & \gate{{\rm R}_z} & \\
\lstick{electronic $e$} && & & & & \gate[2]{{\rm UCR}_z} & \gate{{\rm R}_z} & \\
\lstick{torsion $\theta$} & \qwbundle{m_{\theta}}& & \gate{\rm QFT} & \gate{e^{-i\tau\hat K}} & \gate{{\rm QFT}^\dagger} & & &
\end{quantikz}}
\end{minipage}\\[10pt]
\begin{minipage}[t]{0.86\textwidth}
\centering
(c) Circle embedding\\[4pt]
\resizebox{\linewidth}{!}{%
\begin{quantikz}[row sep=0.22cm,column sep=0.15cm,classical gap=0.02cm]
\lstick{electronic $e$} & & & \gate{{\rm R}_y(\beta)} & \ctrl{3} & \gate{{\rm R}_y(-\beta)} & & \ctrl{1} & \ctrl{1} & \ctrl{2} & \gate{{\rm R}_z} & & & \\
\setwiretype{b}\lstick{torsion $+$} & \gate{\rm K} & \gate[2]{\rm CK} & & & & \gate[2]{\rm BS} & \gate{\rm CD} & \gate{\rm CP} & & \gate{\rm P} & \gate{\rm D} & \gate[2]{{\rm BS}^\dagger} & \\
\setwiretype{b}\lstick{torsion $-$} & \gate{\rm K} & & & & & & & & \gate{\rm CP} & \gate{\rm P} & & & \\
\setwiretype{b}\lstick{vibration $x$} & \gate{\rm D} & & & \gate{\rm CD} & & & & & & \gate{\rm R} & & &
\end{quantikz}}
\end{minipage}
\caption{One time step of Eq.~\eqref{eq:fullstrang} as a circuit, with the kinetic and coupling blocks first and the potential block second. Thin and thick lines represent qubits and qumodes, respectively, with labels indicating their degree of freedom.
A dot marks the qubit whose $\hat\sigma_z$ conditions the gate. Gates follow Eqs.~\eqref{eq:cdconvention} to~\eqref{eq:circleelements}.
(a) Wrapped encoding: three physical lines and one ancilla. The kinetic block is the momentum quadratic phase ${\rm P}_p$, the coupling block is Eq.~\eqref{eq:couplingcircuit}, and the potential block contains the vibration rotation ${\rm R}$, the qubit rotation ${\rm R}_z$, and the two conditional cosine gates of Eq.~\eqref{eq:trigelements}. On $k$ qumodes the conditional displacements of the cosine gates are repeated on each mode and ${\rm P}_p$ splits into the one- and two-mode phases of Eq.~\eqref{eq:resource:gkpkinetic}.
(b) All-qubit register of Sec.~\ref{sec:resource:qubits} with eight torsion and four vibration qubits. The quantum Fourier transform (QFT) brings the torsion to momentum for the kinetic phases, the basis change $\hat B_x$ brings the vibration to position for the coupling block and the uniformly controlled rotation ${\rm UCR}_z$ applies both cosine phases.
(c) Circle embedding on its native pair: four physical lines and no ancilla. Kerr and cross-Kerr phases give the kinetic block of Eq.~\eqref{eq:ringkineticcircuit}. The passive transformation BS maps the circular modes $\pm$ to the Cartesian modes $0$ and $1$, which the two torsion lines carry until ${\rm BS}^\dagger$ maps them back. There, the completed potential acts through the gates of Eq.~\eqref{eq:circleelements} and its Gaussian scalar part.}
\label{fig:circuits}
\end{figure*}
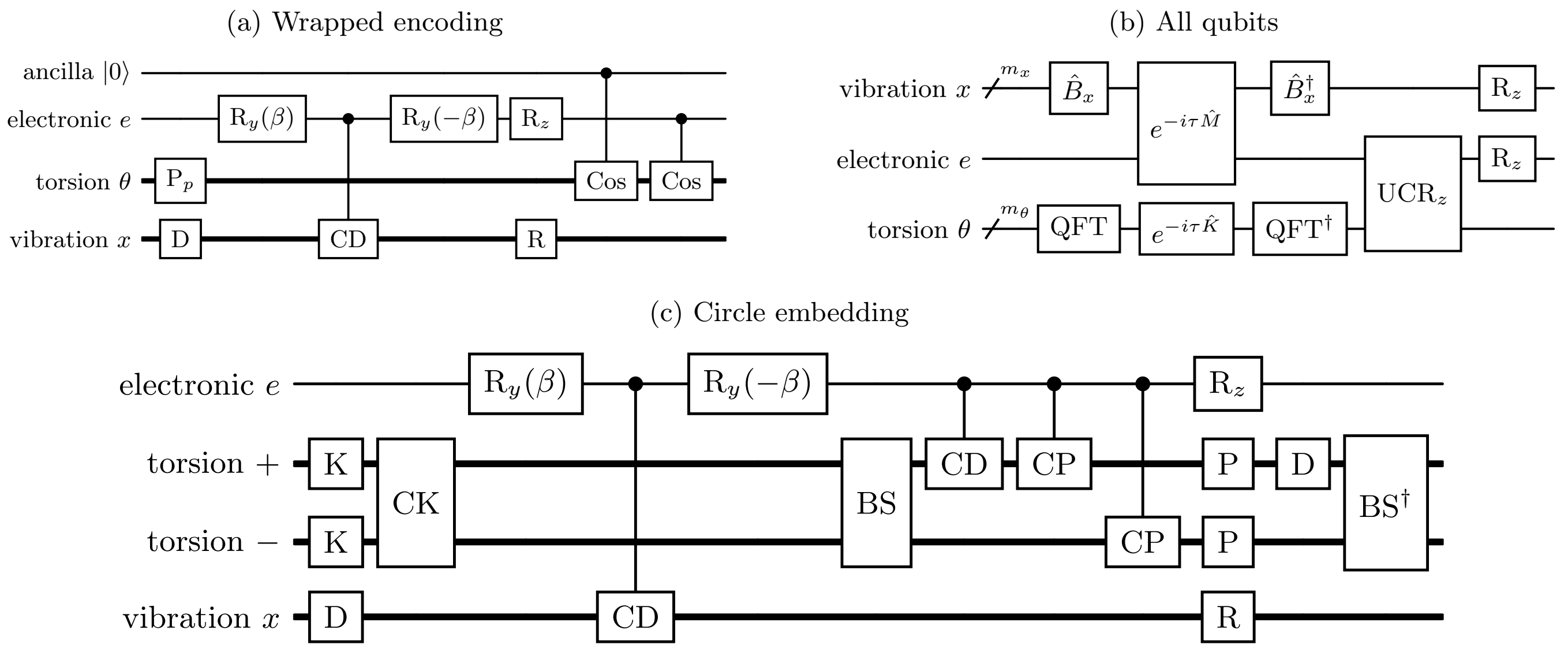

To expose the required blocks, write $\hat{H}=\hat{K}+\hat{D}+\hat{M}$, where
\begin{align}
 \hat{K}&=\frac{G}{2}\hat{L}^2,\nonumber\\
 \hat{D}&=\omega(\hat{n}_x+\one/2)+E_1\frac{\one-\hat{\sigma}_z}{2}
 +(c_0\one+c_z\hat{\sigma}_z)(\one-\cos\hat{\theta}),\nonumber\\
 \hat{M}&=\hat{x}\left(\frac\kappa2\one-\frac\kappa2\hat{\sigma}_z
                                  +\lambda\hat{\sigma}_x\right),
 \label{eq:gateblocks}
\end{align}
with $c_0=(W_0-W_1)/4$ and $c_z=(W_0+W_1)/4$. The kinetic and coupling blocks act on different factors and commute, so a symmetric second-order step is
\begin{equation}
 \hat{S}_2(\Delta t)=e^{-i\hat{K}\Delta t/2}e^{-i\hat{M}\Delta t/2}
              e^{-i\hat{D}\Delta t}e^{-i\hat{M}\Delta t/2}
              e^{-i\hat{K}\Delta t/2}.
 \label{eq:fullstrang}
\end{equation}
The half steps of $\hat{K}+\hat{M}$ merge between consecutive steps. We draw one step for each register in Fig.~\ref{fig:circuits} and identify its gates block by block, with $\tau$ the duration of each factor.

Two gates of $e^{-i\tau\hat{D}}$ are common to every register. The harmonic term is the qumode rotation ${\rm R}(-\omega\tau)$ of the vibration and the constant part is a qubit rotation ${\rm R}_z$, up to a global phase. Its angle is $\tau(E_1-2c_z)$ in the wrapped and qubit registers and $\tau(E_1-2c_z+c_0)$ for the circle, whose completed potentials carry part of the constant. The coupling block is exactly reducible to linear operations. With $v=\sqrt{\lambda^2+(\kappa/2)^2}$, $\sin\beta=\lambda/v$ and $\cos\beta=-\kappa/(2v)$,
\begin{equation}
 e^{-i\tau\hat{M}}={\rm D}\!\left(-\frac{i\tau\kappa}{2\sqrt2}\right)
 {\rm R}_y(-\beta)\,{\rm CD}\!\left(-\frac{i\tau v}{\sqrt2}\right)
 {\rm R}_y(\beta),
 \label{eq:couplingcircuit}
\end{equation}
one displacement of the vibration, one conditional displacement and two qubit rotations in every hybrid register. The remaining factors are the cosine phases $e^{i\tau c_0\cos\hat{\theta}}$ and $e^{i\tau c_z\hat{\sigma}_z\cos\hat{\theta}}$ and the kinetic factor $e^{-i\tau\hat{K}}$, which depend on the encoding.

In the wrapped encoding $\cos\hat{\theta}\to\cos(s\hat{q})$, so the two cosine phases are conditional cosine gates,
\begin{equation}
 e^{i\tau c_z\hat{\sigma}_z\cos\hat{\theta}}={\rm Cos}(-\tau c_z)_{e},
 \qquad
 e^{i\tau c_0\cos\hat{\theta}}={\rm Cos}(-\tau c_0)_{a},
 \label{eq:trigelements}
\end{equation}
where the subscript names the gate qubit. The electronic qubit $e$ provides $\hat{\sigma}_z$ for the first. An ancilla $a$ prepared in $\ket{0}$ provides it for the second, which then acts as a scalar phase and leaves the ancilla in $\ket{0}$ up to the synthesis error, so one ancilla serves every step.
The construction with a single gate qubit~\cite{Chalermpusitarak:2025cod,Rainaldi:2026wpt} realizes ${\rm Cos}(\gamma)$ as a sequence of $n_{\rm c}$ first-order steps, with $2n_{\rm c}+1$ conditional displacements and $2n_{\rm c}$ rotations of the gate qubit. Already the lowest order $n_{\rm c}=1$, three conditional displacements and two rotations, reproduces the gate up to $\mathcal{O}(\gamma^2)$, and the sequence converges to the exact gate as $n_{\rm c}$ grows. We count gates at $n_{\rm c}=1$. The kinetic factor is a quadratic phase in momentum, $e^{-i\tau G\hat{p}^2/(2s^2)}={\rm R}(\pi/2){\rm P}(-\tau G/s^2){\rm R}(-\pi/2)$,
written ${\rm P}_p$ in Fig.~\ref{fig:circuits}. 

For a collective coordinate every conditional displacement factorizes exactly
\begin{equation}
 e^{-i\beta\hat{\sigma}_z\hat{X}}
 =\prod_{j=1}^k
 {\rm CD}\!\left(-\frac{i\beta}{\sqrt{2k}}\right)_{\hat{q}_j},
 \label{eq:resource:displacements}
\end{equation}
where the subscript names the mode acted on. A construction with $M_{\rm CD}$ displacements therefore uses $kM_{\rm CD}$ after distribution, with the same approximation order. The kinetic block also factorizes,
\begin{equation}
 \frac{G}{2s^2}\hat{P}^2=
 \frac{G}{2ks^2}\left(\sum_j\hat{p}_j^2+2\sum_{i<j}\hat{p}_i\hat{p}_j\right),
 \label{eq:resource:gkpkinetic}
\end{equation}
into $k$ one-mode and $\binom{k}{2}$ two-mode commuting Gaussian phases. The kinetic counterterm of Eq.~\eqref{eq:momentumflatten} replaces this phase by $e^{-i\tau Gf_3(\hat{L}_{\mathbb R})^2/2}$, with
\begin{equation}
 f_3(\hat{L}_{\mathbb R})^2=\hat{L}_{\mathbb R}^2
 -\frac{\hat{L}_{\mathbb R}\sin(2\pi\hat{L}_{\mathbb R})}{\pi}
 +\frac{\one-\cos(4\pi\hat{L}_{\mathbb R})}{8\pi^2}.
 \label{eq:resource:f3squared}
\end{equation}
The first term is the Gaussian phase above and the last is a cosine gate on the momentum quadrature. Only the middle term mixes polynomial and trigonometric functions. It can be synthesized from the same native gates by universality~\cite{Lloyd:1998jk,Rainaldi:2025ymn}. The simplest construction is a group commutator of a conditional displacement and a conditional sine gate along the momentum quadrature, which reproduces it to leading order in the gate angles. We do not include these gates in the counts.

For the circle in its two circular modes the kinetic unitary is exactly
\begin{equation}
 e^{-i\tau G\hat{J}^2/2}
 ={\rm K}\!\left(-\frac{\tau G}{2}\right)_{+}
 {\rm K}\!\left(-\frac{\tau G}{2}\right)_{-}
 {\rm CK}(\tau G)_{+-},
 \label{eq:ringkineticcircuit}
\end{equation}
 On $2k$ distributed modes the total angular momentum $\sum_j(\hat{n}_{j,+}-\hat{n}_{j,-})$ acts as $\hat{J}$ on the encoded sector, so its square requires $2k$ Kerr and $\binom{2k}{2}$ cross-Kerr phases. The completed potential of Eq.~\eqref{eq:quadraticcosine} acts on the Cartesian modes, which the passive transformation ${\rm BS}={\rm R}(-\pi/2)_-{\rm BS}(\pi/4)$, a 50:50 beam splitter followed by a quarter rotation of one mode, obtains from the circular ones by inverting Eq.~\eqref{eq:schwinger}. With each diabat completed about its own minimum, the potential block is $\frac{\hat{q}_0^2+\hat{q}_1^2}{2R^2}(c_z\one+c_0\hat{\sigma}_z)-\frac{\hat{q}_0}{R}(c_0\one+c_z\hat{\sigma}_z)$ up to constants. Its $\hat{\sigma}_z$ part gives
\begin{equation}
 {\rm CD}\!\left(\frac{i\tau c_z}{\sqrt2R}\right)_{0}
 {\rm CP}\!\left(-\frac{\tau c_0}{R^2}\right)_{0}
 {\rm CP}\!\left(-\frac{\tau c_0}{R^2}\right)_{1},
 \label{eq:circleelements}
\end{equation}
while its scalar part gives the Gaussian gates ${\rm D}(i\tau c_0/(\sqrt2R))_0{\rm P}(-\tau c_z/R^2)_0{\rm P}(-\tau c_z/R^2)_1$ and the constants join ${\rm R}_z$. The radial counterterm of Eq.~\eqref{eq:boundedflatten} adds cubic and quartic position phases, and polarization identities reduce them to one-mode phases in Gaussian-transformed coordinates.

Per step of Eq.~\eqref{eq:fullstrang}, with the half steps of consecutive steps merged, these constructions give the following counts of qubit-qumode operations. The wrapped encoding on one qumode uses seven conditional displacements, one in the coupling block and three in each of the two cosine gates, with one Gaussian shear and one ancilla qubit. On $k$ qumodes the cosine displacements are repeated on each mode, which gives $6k+1$ conditional displacements and $\binom{k}{2}$ two-mode Gaussian phases, 73 and 66 on 12 qumodes.
The circle on its native pair uses four conditioned operations and one cross-Kerr phase with the completed potential, or eleven with the counterterm. Its potential acts on the Cartesian modes and its kinetic phases on the circular modes, so each step adds two passive transformations between them. The counterterm further adds the Gaussian coordinate changes of its polarization. Single-mode non-Gaussian phases are not counted. These are the two Kerr phases of the circle and, with the counterterm, the seven unconditioned cubic and quartic phases of its scalar part. Conditional displacements are demonstrated primitives, while conditioned quadratic, cubic and quartic phases require a platform-dependent synthesis. Each counted operation is compiled into native gates with a sequence length set by the platform and the target precision.

The two compact encodings offer complementary experimental tradeoffs. With its counterterm, the circle encoding requires only two qumodes and modest squeezing, making it well suited for platforms with native Kerr and higher-order polynomial interactions. Conversely, the GKP encoding is built exclusively from conditional displacements and the demonstrated trigonometric gates~\cite{Rainaldi:2026wpt,McGarry:2026mkk}, at the cost of distributing the unwrapped angular spread across many qumodes.

\subsection{All-qubit construction}
\label{sec:resource:qubits}

A fair comparison with qubits should retain the structure of the molecular Hamiltonian, since encoding and synthesis choices can substantially change the cost of bosonic and vibronic problems \cite{Sawaya:2020bky,Nourse:2025qus,Moeed:2025wtt}. We follow the split-operator approach to chemical dynamics~\cite{Zalka:1996st,Kassal:2008rnk}, with a Fourier grid of $d_\theta=2^{m_\theta}$ points for the torsion and $d_x=2^{m_x}$ Fock levels for the coupling vibration.

The kinetic phase is quadratic in the binary angular momentum and costs $\binom{m_\theta}{2}$ controlled phases. An exact quantum Fourier transform pair, QFT and ${\rm QFT}^\dagger$, between momentum and angle costs $m_\theta(m_\theta-1)$ more \cite{Barenco:1996hq}, with two CNOTs per controlled phase. On the angular grid the cosine factorizes as $\cos\hat{\theta}=\hat{Z}_{\rm t}\otimes\hat{F}$, because a shift by $\pi$ flips the top bit and the sign. A uniformly controlled rotation ${\rm UCR}_z$ on the top bit then implements the scalar phase with $d_\theta/2$ CNOTs \cite{Mottonen:2004vly}. A second one between two CNOTs from the electronic qubit gives the electronic phase, so the cosine block costs $d_\theta+2$ CNOTs. The coupling block is diagonal in the eigenbasis of $\hat{x}$ and costs $2d_x$ CNOTs, together with a basis change $\hat{B}_x$ and its inverse. The basis changes recur at every step because the harmonic term is diagonal in the Fock basis instead. That term is linear in the binary vibrational number and the constant part acts on the electronic qubit, so both are one-qubit rotations ${\rm R}_z$. For $r$ steps, with adjacent half steps merged,
\begin{align}
 C_{\rm circuit}\le{}&r\left[d_\theta+2+2d_x
 +3m_\theta(m_\theta-1)+2C(\hat{B}_x)\right]\nonumber\\
 &+3m_\theta(m_\theta-1)+2d_x+2C(\hat{B}_x).
 \label{eq:fullqubitbudget}
\end{align}
In our case, eight torsional qubits suffice to hold the angular momenta of the reference calculations and four qubits hold 16 vibrational levels, so the register has thirteen qubits including the electronic one.
Therefore, the bound in Eq.~\eqref{eq:fullqubitbudget} with $(m_\theta,m_x)=(8,4)$ gives $658r+400$ CNOTs. Merging diagonal operations or arithmetic phase evaluation can lower this baseline, while routing on hardware with limited connectivity raises it.

One step thus costs 658 CNOTs on qubits, or 1058 for an isolated step. The circle uses four conditioned operations and one cross-Kerr phase per step, or eleven with its counterterm. The wrapped encoding uses seven conditional displacements on a single qumode and, distributed over the twelve qumodes of Fig.~\ref{fig:multimode}, 73 conditional displacements and the 66 two-mode Gaussian phases of Eq.~\eqref{eq:resource:gkpkinetic}.

The two counts rest on very different assumptions. The qubit count assumes error-free logical gates and omits rotation synthesis, routing and error correction, and other qubit encodings might prove more suitable for realistic hardware constraints~\cite{Sawaya:2020bky}. On the other hand, the qumode count treats conditioned polynomial, Kerr and cross-Kerr phases as single gates.

The more robust difference is how the two costs grow. On qubits, the gate count is directly related to the cutoff. The cosine block grows linearly with the number of grid points $d_\theta$, a general basis change of the vibration grows quadratically with $d_x$ and a generic function of several coordinates grows with the product of their grid sizes.
Arithmetic evaluation of the phases reduces this growth at the price of ancillas and a larger constant. For qumodes, the gate parameters do not depend on the cutoff, so a larger Fock range changes only the hardware requirement but not the circuit. When the potential is a sum of few-body terms of bounded polynomial and trigonometric degree, each additional nuclear coordinate adds one qumode, each bilinear coupling adds one Gaussian gate and polynomial terms of fixed degree add a number of gates polynomial in the number of modes, each compiled with a sequence whose length is set by the degree and the target precision. The kinetic, harmonic and coupling blocks are native or exact for qumodes, while the qubit register pays for them through Fourier transforms, multiplexed rotations and basis changes at every step.
Calibrated gate durations and errors on a given platform turn these counts into run times, a natural next step.

%% file: sections/conclusions.tex
\section{Conclusions and outlook}
\label{sec:conclusions}

Quantum simulations of photochemical reactions have largely focused on local expansions around equilibrium geometries, leaving large-amplitude reactive coordinates, such as molecular torsions, outside of their natural scope. Photoisomerization reactions strongly depend on both the global anharmonicity of the potential as well as the global topology of the nuclear configuration space. As a result, treating the reaction coordinate via a local vibronic expansion, a noncompact trigonometric expansion on the line, or on the physical circle leads to qualitatively different dynamics. In this work, we have developed a hybrid qubit-qumode framework that maps each molecular degree of freedom to its native hardware counterpart: electronic states to qubits, vibrations to quantum oscillators and the periodic torsion to a compactly encoded bosonic register. 

The two-mode Hahn-Stock model of retinal photoisomerization makes this distinction especially clear. While linear vibronic approximations can trap the wavepacket near the cis well, restoring the full trigonometric potential allows the wavepacket to accelerate towards the conical-intersection region and branch into the two adiabatic surfaces. Over early timescales, the unwrapped line provides a useful approximation, as the wavepacket explores only the first angular period, but at later times counter-propagating branches that follow different winding paths recombine and interfere on the circle, while on the unwrapped line they remain separated. Under the same potential, the compact topology alters the late-time electronic populations, return probabilities and adiabatic cis/trans branching fractions by $10$--$15\%$ after the first angular period.

Extending this analysis to the more complicated three-mode PSB3 model shows how the physics changes as we increase the number of nuclear coordinates. Compactness affects the product fraction less than in Hahn-Stock, but the late return dynamics remains sensitive to it, so the accuracy of the noncompact approximation strongly depends on the observable of interest. Correlations with the other nuclear coordinates can change how the winding paths overlap and interfere, showing the need for a diverse and complementary set of molecular observables to probe the dynamical role of global nuclear geometry.

To realize the compact torsion on naturally noncompact bosonic hardware, we analyzed and benchmarked two compact qumode encodings: a wrapped Gottesman-Kitaev-Preskill stabilizer code and a two-qumode constrained-circle embedding. Both split the Hilbert space into conserved sectors, labeled by the fractional momentum for the GKP comb and by the radius for the circle. On realistic hardware the state spreads over a narrow distribution of these sectors, which gradually dephases the encoded dynamics. We introduced two independent and scalable methods that reduce the hardware requirements of both encodings without changing the simulated dynamics. Distributing an encoded coordinate across several qumodes lowers the boson occupation of each mode, at the price of more couplings during the time evolution. Hamiltonian counterterms that leave the physical sector unchanged reduce the dependence of the encoded dynamics on the sector label, which relaxes the required squeezing and occupation.

Compiled into elementary gates, the qumode constructions are compact. The kinetic, harmonic and vibronic coupling terms are native or exact on qumodes and their gate count does not depend on the Fock cutoff, while a qubit register pays for the same terms with a gate count that grows with the number of grid points. This structure favors qumodes for potentials built from few-body terms of bounded polynomial and trigonometric degree, the structure of the vibronic coupling models widely used in photochemistry and of the torsional harmonics of PSB3.

Recent advances in trapped-ion and superconducting platforms, including trigonometric gates and conditional displacements on the qumodes, make these photoisomerization simulations promising targets for near-term experiments. A dedicated hardware study will refine these estimates with platform-specific gate errors and noise. A natural first hardware milestone is to prepare a localized cis wavepacket, follow its nonadiabatic evolution through a conical intersection and resolve the changes produced by the full potential and the compact encoding. The electronic population alone carries this signal, so a direct qubit readout is enough. The molecular benchmarks developed here set the accuracy needed to distinguish these effects. The circle embedding with its radial counterterms runs on its native pair of qumodes with modest squeezing, which makes it a first possible target for such an experiment.
The wrapped encoding uses only demonstrated gate types and distributes its occupation over more qumodes as the torsion completes full turns.

Progress toward utility in quantum chemistry will require extending this connection between molecular observables and physical resources. Additional reactive coordinates, couplings to an environment and potentials informed by electronic-structure calculations are natural next steps. They will allow studies of how nuclear correlations and environmental relaxation alter photochemical branching and recurrences. Comparison with established classical dynamics methods can identify which regimes benefit most from retaining the coupled quantum motion. By treating compact reaction coordinates within the same qubit-qumode description as electronic transitions and noncompact vibrations, the present framework provides the compact coordinates that quantum simulations of larger photochemical systems require.

\begin{table*}[t!]
\caption{PSB3 model parameters.}
\label{tab:psb3parameters}
\centering\normalsize
\setlength{\tabcolsep}{5pt}
\renewcommand{\arraystretch}{1.15}
\begin{tabular*}{\textwidth}{@{\extracolsep{\fill}}lrl@{\hspace{2.5em}}lrl@{}}
\toprule
Parameter & Value & Unit & Parameter & Value & Unit \\
\cmidrule(r{1em}){1-3}\cmidrule(l{1em}){4-6}
$D$ & $2000$ & $\mathrm{kcal\,mol^{-1}}$ & $c_1$ & $437.068$ & $\mathrm{kcal\,mol^{-1}}\,\mathrm{\text{\AA}^{-2}}$ \\
$r_{\rm TS}$ & $0.025079167$ & $\mathrm{\text{\AA}}$ & $c_2$ & $16.7317$ & $\mathrm{kcal\,mol^{-1}}\,\mathrm{\text{\AA}^{-1}}$ \\
$r_{\rm min}$ & $0.0912615$ & $\mathrm{\text{\AA}}$ & $c_3$ & $7.35468$ & $\mathrm{kcal\,mol^{-1}}$ \\
$d_1$ & $\sqrt{3733.5/(2D)}$ & $\mathrm{\text{\AA}^{-1}}$ & $c_4$ & $88.517$ & $\mathrm{kcal\,mol^{-1}}$ \\
$d_4$ & $\sqrt{1097.19/(2D)}$ & $\mathrm{\text{\AA}^{-1}}$ & $c_5$ & $5.95221$ & $1$ \\
$d_2$ & $54.633$ & $\mathrm{kcal\,mol^{-1}}$ & $h_1$ & $155.749$ & $\mathrm{kcal\,mol^{-1}}$ \\
$d_3$ & $3.8008$ & $\mathrm{kcal\,mol^{-1}}$ & $k_1$ & $24.0358$ & $\mathrm{kcal\,mol^{-1}}$ \\
\bottomrule
\end{tabular*}

\vspace{0.5em}
\begin{tabular*}{\textwidth}{@{\extracolsep{\fill}}lrlrlr@{}}
\toprule
\multicolumn{6}{@{}l}{Kinetic coefficients (atomic units)} \\
\midrule
$G_{rr}$ & $7.981\times10^{-5}$ &
$G_{\theta\theta}$ & $2.599\times10^{-5}$ &
$G_{\phi\phi}$ & $40.375\times10^{-5}$ \\
\bottomrule
\end{tabular*}

\end{table*}

\begin{acknowledgments}
We thank Zohreh Davoudi, Matt Grau, Norbert Linke, and Peter Love for helpful discussions. F.H. is supported by the Austrian Science Fund (FWF), project 10.55776/J4854. T.R. and F.R. are supported by the U.S. Department of Energy, Office of Science, Office of Nuclear Physics, Early Career Program under contract No. DE-SC0025881.
This research used resources of the National Energy Research Scientific Computing Center (NERSC), a U.S. Department of Energy Office of Science User Facility, under NERSC award NP-ERCAP0037787.
We thank Stony Brook Research Computing and Cyberinfrastructure and the Institute for Advanced Computational Science at Stony Brook University for access to SeaWulf. The system was funded by National Science Foundation award 1531492 and Major Research Instrumentation award 2215987, with matching funds from Empire State Development's Division of Science, Technology and Innovation (NYSTAR) under contract C210148.
\end{acknowledgments}

%% file: sections/appendices.tex
\appendix

\section{Numerical conventions and convergence}
\label{app:numerical}

All reference calculations represent each degree of freedom in a basis
with a cutoff, a Fourier basis for the torsion and a coordinate or
harmonic-oscillator basis for the coupling vibration. Every cutoff is
increased until the reported observables are converged well below the
simulation tolerance. On position grids we propagate with the
split-operator method. The kinetic factor is applied exactly in
momentum space by fast Fourier transforms and the potential factor, a
$2\times2$ matrix in the electronic states at each grid point, is
exponentiated exactly point by point. In a Fock or angular-momentum
basis we instead use a Chebyshev expansion of the propagator. We
compare trajectories at identical sample times, since $P_1$ and the
return probability oscillate on the $1.7$ fs electronic period and
interpolation between samples would add spurious differences.

The noncompact line is represented by joining $K$ copies of the angular
cell into one periodic box of circumference $2\pi K$, with $K$ large
enough that the outer cells remain empty over the reported times. Two
independent implementations agree far below the tolerance. In the
basin readouts, boundary nodes receive half weight.

For the circle encoding the radial measure is $\rho\,d\rho$ and
finite-width reference dynamics are assembled from radial sectors. The
reference ansatz $f_A(\rho)\psi_{\rm mol}(\theta,x)$ gives every
angular harmonic the same radial profile. A physical planar state
vanishes smoothly at the origin, which this ansatz satisfies to
numerical precision because the prepared annulus carries no resolvable
probability there.

The fiber tolerances used in Sec.~\ref{sec:resource:occupation} follow from
exact propagation of single fibers of the circle and the comb,
Eqs.~\eqref{eq:radialsector} and~\eqref{eq:blochsector}, with and
without the counterterms of Eqs.~\eqref{eq:boundedflatten}
and~\eqref{eq:momentumflatten}, with each circle diabat completed about
its own minimum. A Gaussian fiber distribution of width
$\sigma$ weights the recorded trajectories, wrapped onto
$-1/2\le\nu<1/2$ for the comb. The tolerance is the
largest $\sigma$ for which $P_1$ stays within the target of its
physical-fiber value, $\sigma_\nu\le0.082$ and $0.22$ for the comb
without and with the kinetic counterterm and $1.1\%$ and $5.5\%$ for
the circle without and with the radial counterterm. The same propagations also record $Q_-$, $Y_-$ and
the return. Holding these within the target as well narrows the
deterministic tolerances to $\sigma_\nu\le0.068$ and $0.19$ for the
comb and $0.12\%$ and $4.3\%$ for the circle. The most sensitive
observable is $Y_-$ for the comb and the coherent return for the bare
circle.

The finite-cutoff benchmarks propagate the projected Hamiltonian
$\mathcal{P}\hat{H}\mathcal{P}$, with $\mathcal{P}$ the projector onto
the retained Fock states, after projecting and normalizing the initial
state. Projecting an operator and applying a function to the projected
operator are different operations. With $\mathcal{P}_N$ the projector
onto $\ket{0},\dots,\ket{N-1}$ of one mode, for example,
\begin{equation}
 \mathcal{P}_N\hat{p}^2\mathcal{P}_N=(\mathcal{P}_N\hat{p}\mathcal{P}_N)^2+
           \frac N2\ket{N-1}\bra{N-1}.
 \label{eq:appendix:p2projection}
\end{equation}
In all Fock-basis calculations we truncate the Hamiltonian itself. We
compute its matrix elements in the full Fock basis and keep the block
of the retained levels. The circle benchmark propagates the projected
polynomial Hamiltonian with independent cutoffs on $n_+$ and $n_-$.
These runs use the linear substitution $\cos\hat{\theta}\to\hat{q}_0/R$,
which coincides with the completed form of
Eq.~\eqref{eq:quadraticcosine} on the constraint surface. We
check the completed form fiber by fiber. At $A=192$ these
fiber propagations give a finite-width error in $P_1$ of $0.0030$ for
the linear substitution and $0.0017$ for the completion.

The electronic populations and return probabilities are evaluated in
the evolution basis, while the adiabatic and basin observables require
a coordinate-space quadrature. If that readout misses probability
$\delta$, with captured lower-surface and trans probabilities $L_-$ and
$T_-$, then
\begin{equation}
 \frac{T_-}{L_-+\delta}\le Y_-\le
 \frac{T_-+\delta}{L_-+\delta}.
 \label{eq:appendix:readoutbound}
\end{equation}
These intervals give the bound on the circle error in $Y_-$ used in
Sec.~\ref{sec:finiteresults}.

\section{PSB3 parameters}
\label{app:psb3}

Table~\ref{tab:psb3parameters} lists the parameters of the PSB3 model of
Sec.~\ref{sec:psb3:model}. The radial functions of
Eq.~\eqref{eq:psb3radialfunctions} take the BLA coordinate $r_{\rm A}$ in
angstroms and the potential energies are converted from kcal/mol to
Hartree by the factor $1/627.5094740631$ before they are combined with
the kinetic operator.